\documentclass[11pt]{article}
\usepackage{amsmath,amssymb,graphicx}

\newcommand{\ba}{\begin{alignat}{3}}

\begin{document}

\begin{titlepage}
\begin{flushright}
%   IU-MSTP/ \\
%   14 November 2012
\end{flushright}
\begin{center}
  \vspace{4cm}
  {\bf \Large Boosted Quantum Black Hole and Black String in M-theory, \\[0.2cm]
  and Quantum Correction to Gregory-Laflamme Instability}
  \\  \vspace{2cm}
  Yoshifumi Hyakutake%\footnote{E-mail: hyaku@mx.ibaraki.ac.jp}
   \\ \vspace{1cm}
   {\it Faculty of Science, Ibaraki University \\
   Bunkyo 2-1-1, Mito, Ibaraki 310-8512, Japan}
\end{center}

\vspace{2cm}
\begin{abstract}
We take into account higher derivative $R^4$ corrections in M-theory and 
construct quantum black hole and black string solutions in 11 dimensions
up to the next leading order.
The quantum black string is stretching along the 11th direction and
the Gregory-Laflamme instability is examined at the quantum level.
Thermodynamics of the boosted quantum black hole and black string are also discussed.
Especially we take the near horizon limit of the quantum black string and investigate
its instability quantitatively.
\end{abstract}
\end{titlepage}

\setlength{\baselineskip}{0.65cm}

%%%%%%%%%%%%%%%%%%%%%%%%%%%%%%%%%%%%%%%%%%%%%%%%%%%%%%%%%%%%%%%%%%%%%%%%%%%%%%%%%%%%%%%%%%%%%%%
%%%%%%%%%%%%%%%%%%%%%%%%%%%%%%%%%%%%%%%%%%%%%%%%%%%%%%%%%%%%%%%%%%%%%%%%%%%%%%%%%%%%%%%%%%%%%%%
\section{Introduction}

One of challenging problems in theoretical physics is to reveal the nature of the quantum gravity.
Especially the superstring theory is a good candidate for the quantum gravity in which 
properties of black holes have been studied enormously.
In this paper we will study the quantum aspects of the black hole
and the black string in superstring theory and M-theory.

The type IIA superstring theory is perturbatively defined in 10 dimensions, and its low energy limit
is described by the type IIA supergravity. 
This theory contains black hole solution which possesses SO(9) rotation symmetry.
Interestingly the strong coupling limit of the type IIA superstring theory is believed to be described by 
11 dimensional M-theory\cite{Townsend:1995kk,Witten:1995ex}. 
The 11th direction is compactified on a circle
and its radius is proportional to the string coupling constant.
The low energy limit of the M-theory is approximated by the 11 dimensional supergravity\cite{Cremmer:1978km},
and it contains black hole solution which has SO(10) rotation symmetry.
Since the type IIA supergravity is obtained by Kaluza-Klein reduction of the 11 dimensional supergravity\cite{Huq:1983im},
the 10 dimensional black hole with SO(9) symmetry is identified with a 11 dimensional black string 
which is stretching along the 11th direction.

The black string is stable against perturbation when the radius of the compactified circle is small.
However, as we enlarge the circle, it becomes unstable at the critical radius and starts to transit 
to the black hole\cite{Gregory:1993vy}. 
This phenomenon is called Gregory-Laflamme instability and has been studied since 1993\cite{Gregory:1994bj}.
In 2000, the fate of this instability is analyzed and the existence of non-uniform black
string is discussed in ref.~\cite{Horowitz:2001cz}.
After this work, there appeared several analytic studies and numerical simulations
in order to resolve the final state of the unstable black string\cite{Gubser:2001ac}-\cite{Lehner:2010pn}. 
These show that, beyond the critical radius, the black string transit into 
the non-uniform string and finally pinch off.
It is also revealed that the fate of the unstable black string depends on the dimension of the spacetime
and boost parameter\cite{Sorkin:2004qq,Hovdebo:2006jy}.
Comprehensive reviews on the Gregory-Laflamme instabilities are collected in the book \cite{Horowitz:2012nnc}.

In this paper we focus on the black string solution in 11 dimensions and examine its instability.
Especially, since the superstring theory and M-theory contain quantum corrections to the 
supergravity\cite{Gross:1986iv}-\cite{Grisaru:1986vi}, 
it is natural to study quantum corrections to the Gregory-Laflamme instability.
This is one of the main purpose of this paper, and we take into account $R^4$ corrections to the 
11 dimensional supergravity\cite{Tseytlin:2000sf,Becker:2001pm}. 
These corrections corresponds to the leading 1-loop corrections in the type IIA superstring theory,
and supersymmetric completions of these terms are discussed in refs.~\cite{deRoo:1992sm}-\cite{Hyakutake:2007sm}.
In this paper we briefly review the structure of the $R^4$ corrections in M-theory,
and solve equations of motion for the configurations of the black hole and the black string.
We will call these solutions quantum black hole and quantum black string, respectively.
See refs.~\cite{Callan:1988hs,Brihaye:2010me,Giacomini:2015dwa} for the related works on this topic.

Another important task of this paper is to consider the Lorentz boost of the quantum black string. 
In the case of classical black string, it is known that its instability highly depends 
on the boost parameter\cite{Hovdebo:2006jy}.
Similarly we will see that the thermodynamics of the quantum black string changes and
the its transition point is also be modified due to the boost parameter. 
We also investigate the near horizon limit of the boosted quantum black string,
which corresponds to that of the nonextremal quantum black 0-brane. 

Organization of this paper is as follows.
In sec.~\ref{sec:BH}, we review the higher derivative corrections in M-theory, and
derive equations of motion. 
We consider the quantum corrections to the black hole and the black string
and discuss the instability from entropic arguments.
In sec.~\ref{sec:boost}, we boost the quantum black hole and black string, 
and examine the Gregory-Laflamme instability.
In sec.~\ref{sec:nearH}, the near horizon limit of the quantum black string is investigated.
Section~\ref{sec:conc} is devoted to the summary of this paper and future works.
Technical calculations on both quantum black hole and black string are explicitly shown 
in appendices \ref{sec:appA} and \ref{sec:appB}.

%%%%%%%%%%%%%%%%%%%%%%%%%%%%%%%%%%%%%%%%%%%%%%%%%%%%%%%%%%%%%%%%%%%%%%%%%%%%%%%%%%%%%%%%%%%%%%%
%%%%%%%%%%%%%%%%%%%%%%%%%%%%%%%%%%%%%%%%%%%%%%%%%%%%%%%%%%%%%%%%%%%%%%%%%%%%%%%%%%%%%%%%%%%%%%%
\section{Quantum Black Hole and Black String in M-theory} \label{sec:BH}

%%%%%%%%%%%%%%%%%%%%%%%%%%%%%%%%%%%%%%%%%%%%%%%%%%%%%%%%%%%%%%%%%%%%%%%%%%%%%%%%%%%%%%%%%%%%%%%
\subsection{Brief Review of Higher Derivative Corrections in M-theory}

M-theory is defined as a strong coupling limit of type IIA superstring theory,
and it is well approximated by 11 dimensional supergravity in the low energy limit.
Classical solutions of 11 dimensional supergravity play important roles to reveal
the structure of the M-theory. In this section, we consider black hole and black string 
solutions in 11 dimensions. These are quite simple and it is possible to discuss 
leading quantum corrections to them.

It is known that the leading correction comes from $R^4$ terms\cite{Tseytlin:2000sf,Becker:2001pm}.
The bosonic part of the M-theory effective action which is relevant to the graviton is given by
\begin{alignat}{3}
  S_{11} &= \frac{1}{2 \kappa_{11}^2} \int d^{11}x \; e \Big\{ R +
  \gamma \Big(t_8 t_8 R^4 - \frac{1}{4!} \epsilon_{11} \epsilon_{11} R^4 \Big) \Big\}
  \notag
  \\
  &= \frac{1}{2 \kappa_{11}^2} \int d^{11}x \; e \Big\{ R +
  24 \gamma \big( R_{abcd} R_{abcd} R_{efgh} R_{efgh}
  - 64 R_{abcd} R_{aefg} R_{bcdh} R_{efgh} \notag
  \\
  &\qquad\qquad\qquad\qquad\qquad\quad
  + 2 R_{abcd} R_{abef} R_{cdgh} R_{efgh}
  + 16 R_{acbd} R_{aebf} R_{cgdh} R_{egfh} \notag
  \\
  &\qquad\qquad\qquad\qquad\qquad\quad
  - 16 R_{abcd} R_{aefg} R_{befh} R_{cdgh}
  - 16 R_{abcd} R_{aefg} R_{bfeh} R_{cdgh} \big) \Big\}, \label{eq:R4}
\end{alignat}
where $a,b,c,\cdots = 0,1,\cdots,10$ are local Lorentz indices.
In this action, the expansion parameter is expressed in terms of the 11 dimensional Planck length as
\begin{alignat}{3}
  \gamma = \frac{\pi^2\ell_p^6}{2^{11} 3^2}. \label{eq:gamma} 
\end{alignat} 
In eq.~(\ref{eq:R4}) pairs of indices are lowered for simplicity, 
but of course they should be contracted by the flat metric.
Note that the above action preserves local supersymmetry, and fermionic terms are given 
in refs.~\cite{Peeters:2000qj}-\cite{Hyakutake:2007sm}.
Note also that $\gamma \sim g_s^2 \ell_s^6$, so if we reduce the effective action (\ref{eq:R4}) 
into 10 dimensions, it corresponds to 1-loop quantum correction to the type IIA supergravity.

By varying the effective action (\ref{eq:R4}) with respect to the vielbein, 
we obtain equations of motion,
\begin{alignat}{3}
  E_{ij} &\equiv R_{ij} \!-\! \frac{1}{2} \eta_{ij} R 
  + \gamma \Big\{ \!\!-\! \frac{1}{2} \eta_{ij} \Big(t_8 t_8 R^4 \!-\! \frac{1}{4!} \epsilon_{11} \epsilon_{11} R^4 \Big)
  \!+\! R_{abci} X^{abc}{}_j \!-\! 2 D_{(a} D_{b)} X^a{}_{ij}{}^b \Big\} = 0, \label{eq:MEOM}
\end{alignat}
up to the linear order of $\gamma$.
Here $D_a$ is a covariant derivative for local Lorentz indices and $X_{abcd}$ is defined as
\begin{alignat}{3}
  X_{abcd} &= \frac{1}{2} \big( X'_{[ab][cd]} + X'_{[cd][ab]} \big), \label{eq:X}
  \\
  X'_{abcd} &= 96 \big(
  R_{abcd} R_{efgh} R_{efgh} \!-\! 16 R_{abce} R_{dfgh} R_{efgh} \!+\! 2 R_{abef} R_{cdgh} R_{efgh} 
  \!+\! 16 R_{aecg} R_{bfdh} R_{efgh} \notag
  \\
  &\qquad\;\;
  -\! 16 R_{abeg} R_{cfeh} R_{dfgh} \!-\! 16 R_{efag} R_{efch} R_{gbhd} 
  \!+\! 8 R_{abef} R_{cegh} R_{dfgh} \big) \notag.
\end{alignat}
It is not obvious but possible to check that $R_{abci} X^{abc}{}_j = R_{abcj} X^{abc}{}_i$,
hence $E_{ij}$ is a symmetric tensor.
The explicit derivation of eq.~(\ref{eq:MEOM}) can be found in ref.~\cite{Hyakutake:2013vwa}.

%%%%%%%%%%%%%%%%%%%%%%%%%%%%%%%%%%%%%%%%%%%%%%%%%%%%%%%%%%%%%%%%%%%%%%%%%%%%%%%%%%%%%%%%%%%%%%%%%%%
\subsection{Quantum Black Hole}
\label{sec:QBH}

First we briefly review Schwarzschild black hole solution in 11 dimensional supergravity.
The metric of the black hole is given by
\begin{alignat}{3}
  &ds_\text{h}^2 = - A dt^2 + A^{-1} dr^2 + r^2 d\Omega_9^2, \qquad 
  A = 1 - \frac{r_\text{h}^8}{r^8},  \label{eq:11BH}
\end{alignat}
and the event horizon $r_\text{horizon}$ is located at $r_\text{horizon} = r_\text{h}$.
From standard calculations, ADM mass and entropy of the black hole are evaluated as
\begin{alignat}{3}
  M_\text{h} = \frac{9 V_{S^9} r_\text{h}^8}{2\kappa_{11}^2}, \qquad
  S_\text{h} = \frac{4\pi V_{S^9} r_\text{h}^9}{2\kappa_{11}^2}.
\end{alignat}
Here $2\kappa_{11}^2 = (2\pi)^8 \ell_p^9$ and $\ell_p$ is the Planck length 
in 11 dimensions\footnote{By using string length $\ell_s$ and string coupling constant $g_s$,
the Planck length is expressed as $\ell_p = \ell_s g_s^{1/3}$}.
$V_{S^9}=\frac{\pi^5}{12}$ is the volume of the 9 dimensional unit sphere.

Next let us take into account the quantum correction to the black hole by solving the eq.~(\ref{eq:MEOM}) 
up to the linear order of $\gamma$.
The leading part of the metric (\ref{eq:11BH}) itself is not a solution of the eq.~(\ref{eq:MEOM}),
so we should relax the ansatz. 
Most general static ansatz with SO(10) rotation symmetry is given by
\begin{alignat}{3}
  ds_\text{h}^2 &= - B_1^{-1} A_1 dt^2 + A_1^{-1} dr^2
  + r^2 d\Omega_9^2, \label{eq:bhansatz}
  \\
  A_1 &= 1 - \frac{r_\text{h}^8}{r^8} + \frac{\gamma}{r_\text{h}^6} a_1 \Big(\frac{r}{r_\text{h}}\Big), \qquad
  B_1 = 1 + \frac{\gamma}{r_\text{h}^6} b_1 \Big(\frac{r}{r_\text{h}}\Big). \notag
\end{alignat}
In order to make the equations of motion simple, we introduce following dimensionless coordinates,
$\tau = \frac{t}{r_\text{h}},\, x = \frac{r}{r_\text{h}}$,
and insert the ansatz (\ref{eq:bhansatz}) into the eq.~(\ref{eq:MEOM}).
Then the equations of motion become
\begin{alignat}{3}
  E_1 &= - x^{39} a_1' - 8 x^{38} a_1 - 9299558400 x^8 + 10492093440 = 0, \notag
  \\
  E_2 &= x^{39} a_1' + 8 x^{38} a_1 + x^{31} (1 - x^8) b_1' - 312729600 x^8 - 879805440 = 0, \label{eq:EOM-h}
  \\
  E_3 &= x^{40} a_1'' + 16 x^{39} a_1' + 56 x^{38} a_1 + x^{32} (1 - x^8) b_1'' 
  - 4 x^{31} (1 + 2 x^8) b_1' \notag
  \\
  &\quad\, 
  + 7192780800 x^8 - 11175183360 = 0, \notag
\end{alignat}
where the prime represents the derivative with respect to $x$.
By solving $E_1 = E_2 = 0$ with requiring asymptotic flatness, we obtain
\begin{alignat}{3}
  a_1(x) &= - \frac{349736448}{x^{38}} + \frac{422707200}{x^{30}} + \frac{c_\text{h}}{x^8}, \notag
  \\
  b_1(x) &= \frac{320409600}{x^{30}}, \label{eq:solab}
\end{alignat}
where $c_\text{h}$ is an integral constant.
From this we see that the quantum corrections become important when $r < r_\text{h}$ or
$1 \ll \frac{\gamma}{r_\text{h}^6}$.
Notice that $c_\text{h}$ can be absorbed by the redefinition of $r_\text{h}$ like
\begin{alignat}{3}
  r_\text{h}^8 - \gamma c_\text{h} r_\text{h}^2 \;\to\; r_\text{h}^8,
\end{alignat}
up to linear order of $\gamma$.
So physical quantities do not depend on $c_\text{h}$ 
at this order\footnote{$r_\text{h}^8 - \gamma c_\text{h} r_\text{h}^2 > 0$ is required
since the mass (\ref{eq:M_h}) should be positive.}.
The remaining equation $E_3 = 0$ is trivially satisfied by inserting eq.~(\ref{eq:solab}).
The plots of $A_1(x)$ are shown in appendix \ref{sec:A1}.

Now we call the metric of eq.~(\ref{eq:bhansatz}) with eq.~(\ref{eq:solab}) the quantum black hole.
Let us investigate the thermodynamics of the quantum black hole.
The event horizon is located at $r_\text{horizon}=r_\text{h} - \frac{\gamma}{8 r_\text{h}^5} a_1(1)$
up to the linear order of $\gamma$, and the temperature is given by
\begin{alignat}{3}
  T_\text{h} &= \frac{1}{4\pi} B_1^{-\frac{1}{2}} \frac{d A_1}{dr} \Big|_{r_\text{horizon}} \notag
  \\
  &= \frac{2}{\pi r_\text{h}} \Big\{ 1 + \gamma
  \Big( \frac{9}{8} a_1(1) + \frac{1}{8} a'_1(1) - \frac{1}{2} b_1(1) \Big) \frac{1}{r_\text{h}^6} \Big\}
  \equiv \frac{2}{\pi} \bar{T}_\text{h}.
  \label{eq:T_h}
\end{alignat}
By solving the above equation inversely, $r_\text{h}$ is expressed as
\begin{alignat}{3}
  r_\text{h} &= \frac{1}{\bar{T}_\text{h}} 
  \Big\{1 + \gamma \Big( \frac{9}{8}a_1(1) + \frac{1}{8}a'_1(1) - \frac{1}{2}b_1(1) \Big) 
  \bar{T}_\text{h}^6 \Big\}, \label{eq:r_h}
\end{alignat}
and from this relation, physical quantities can be expressed in terms of the temperature
up to the linear order of $\gamma$.
For instance, the location of the event horizon is evaluated as
\begin{alignat}{3}
  r_\text{horizon} &= \frac{1}{\bar{T}_\text{h}} 
  \Big\{1 + \gamma \Big( a_1(1) + \frac{1}{8}a'_1(1) - \frac{1}{2}b_1(1) \Big) \bar{T}_\text{h}^6 \Big\} \notag
  \\
  &= \frac{1}{\bar{T}_\text{h}} \big( 1 - 11137920 \gamma \bar{T}_\text{h}^6 \big). \label{eq:r_horizon}
\end{alignat}
This reveals that the position of the event horizon slightly moves inward due to the quantum correction,
and its value does not depend on $c_\text{h}$.

The ADM mass $M_\text{h}$ is calculated as
\begin{alignat}{3}
  \frac{2\kappa_{11}^2}{9 V_{S^9}} M_\text{h} &= r_\text{h}^8 \Big( 1 - \gamma \frac{c_\text{h}}{r_\text{h}^6} \Big) \notag
  \\
  &= \frac{1}{\bar{T}_\text{h}^8} \Big\{ 1 + \gamma \big( 9 a_1(1) + a'_1(1) - 4 b_1(1) - c_\text{h} \big) 
  \bar{T}_\text{h}^6 \Big\}, \notag
  \\
  &= \frac{1}{\bar{T}_\text{h}^8} \big( 1 - 16132608 \gamma \bar{T}_\text{h}^6 \big) 
  \equiv \bar{M}_\text{h}. \label{eq:M_h}
\end{alignat}
Note that although the effective action (\ref{eq:R4}) contains the higher derivative terms,
the expression of the ADM mass formula does not\cite{Hyakutake:2014maa}.
The above correction just enters through $\frac{c_\text{h}}{x^8}$ term in $a_1(x)$ and $r_\text{h}$ in eq.~(\ref{eq:r_h}).
On the other hand, the area law of the entropy is modified by the higher derivative corrections \cite{Wald:1993nt,Iyer:1994ys},
and the black hole entropy $S_\text{h}$ is given by
\begin{alignat}{3}
  \frac{2\kappa_{11}^2}{4\pi V_{S^9}} S_\text{h} &= 
  r_\text{horizon}^9 \Big( 1 - 2 \gamma X_{0101} \Big|_{x=1} \Big) \notag
  \\
  &= \frac{1}{\bar{T}_\text{h}^9} \big( 1 - 12099456 \gamma \bar{T}_\text{h}^6 \big) \notag
  \\[0.2cm]
  &= \bar{M}_\text{h}^{9/8}
  \big( 1 + 6049728 \, \gamma \bar{M}_\text{h}^{-3/4} \big).
  \label{eq:S_h}
\end{alignat}
As explained before, $r_\text{horizon}$, $M_\text{h}$ and $S_\text{h}$ 
are written in terms of $T_\text{h}$, and do not depend on the unknown constant $c_\text{h}$. 
It is easy to see that the first law $dM_\text{h} = T_\text{h} dS_\text{h}$ holds up
to the linear order of $\gamma$.

%%%%%%%%%%%%%%%%%%%%%%%%%%%%%%%%%%%%%%%%%%%%%%%%%%%%%%%%%%%%%%%%%%%%%%%%%%%%%%%%%%%%%%%%%%%%%%%%%%%%%
\subsection{Quantum Black String} \label{sec:QBS}

In this section, we consider a black string solution which is constructed by
aligning 10 dimensional black hole along the 11th direction.
The 11th direction is compactified on a circle and its radius is given by $R_{11}=\ell_s g_s$.
In 11 dimensional supergravity, the black string solution is simply given by
\begin{alignat}{3}
  &ds_\text{s}^2 = - F dt^2 + F^{-1} dr^2 + r^2 d\Omega_8^2 + dz^2, \qquad 
  F = 1 - \frac{r_\text{s}^7}{r^7}. \label{eq:11BS}
\end{alignat}
The ADM mass and the entropy are calculated as
\begin{alignat}{3}
  M_\text{s} &= \frac{8 V_{S^8} r_\text{s}^7}{2\kappa_{10}^2}, \qquad
  S_\text{s} &= \frac{4\pi V_{S^8} r_\text{s}^8}{2\kappa_{10}^2},
\end{alignat}
where $V_{S^8}=\frac{2^5\pi^4}{105}$ is the volume of the 8 dimensional unit sphere,
and $2\kappa_{10}^2 = 2\kappa_{11}^2/(2\pi R_{11})$.
These expressions simply show that the 11 dimensional black string corresponds to
the 10 dimensional black hole after the dimensional reduction.

Now we take into account the quantum corrections to the black string. 
Since the metric does not satisfy the equations of motion (\ref{eq:MEOM}),
we relax the ansatz as follows.
\begin{alignat}{3}
  ds_\text{s}^2 &= - G_1^{-1} F_1 dt^2 + F_1^{-1} dr^2
  + r^2 d\Omega_8^2 + G_2 dz^2, \label{eq:bsansatz}
  \\
  F_1 &= 1 - \frac{r_\text{s}^7}{r^7} + \frac{\gamma}{r_\text{s}^6} f_1 \Big(\frac{r}{r_\text{s}}\Big), \qquad
  G_i = 1 + \frac{\gamma}{r_\text{s}^6} g_i \Big(\frac{r}{r_\text{s}}\Big). \notag
\end{alignat}
This is the most general ansatz which preserves SO(9) rotation symmetry.
In order to make the equations of motion simple, we introduce dimensionless coordinates,
$\tau = \frac{t}{r_\text{h}},\, x = \frac{r}{r_\text{h}},\, y = \frac{z}{r_\text{h}}$,
and insert the ansatz (\ref{eq:bsansatz}) into eq.~(\ref{eq:MEOM}).
Then the equations of motion become
\begin{alignat}{3}
  E_1 &= -16 x^{35} f_1' - 112 x^{34} f_1 + 2 x^{29} ( 1 - x^7 ) g_2'' + x^{28} (9 -16 x^7 ) g_2' \notag
  \\
  &\quad\,
  - 63402393600 x^7 + 71292856320 = 0, \notag
  \\
  E_2 &= 16 x^{35} f_1' + 112 x^{34} f_1 + 16 x^{28} ( 1 - x^7 ) g_1' - x^{28} (9 - 16 x^7 ) g_2' \notag
  \\
  &\quad\,
  - 2159861760 x^7 - 5730600960 = 0, \label{eq:EOM-s}
  \\
  E_3 &= 2 x^{36} f_1'' + 28 x^{35} f_1' + 84 x^{34} f_1 + 2 x^{29} (1 - x^7 ) g_1''
  - 7 x^{28} ( 1 + 2 x^7 ) g_1' \notag
  \\
  &\quad\,
  - 2 x^{29} ( 1 - x^7 ) g_2'' + 14 x^{35} g_2' + 5669637120 x^7 - 8626383360 = 0, \notag
  \\
  E_4 &= 2 x^{36} f_1'' + 32 x^{35} f_1' + 112 x^{34} f_1 + 2 x^{29} ( 1 - x^7 ) g_1''
  - x^{28} ( 5 + 16 x^7 ) g_1' \notag
  \\
  &\quad\,
  - 1062512640 = 0. \notag
\end{alignat}
By solving the equations of motion (\ref{eq:MEOM}) up to the linear order of $\gamma$,
the functions $f_1$ and $g_i (i=1,2)$ are explicitly solved as
\begin{alignat}{3}
  f_1(x) &= - \frac{1208170880}{9x^{34}} 
  + \frac{161405664}{x^{27}} 
  + \frac{5738880}{13 x^{20}}
  + \frac{956480}{x^{13}}
  + \frac{c_\text{s}}{x^7} 
  + \frac{819840}{x^7} I(x), \notag
  \\[0.2cm]%%%%%%%%%%%%%%%%%%%%%%%%%
  g_1(x) &= \frac{1035722240}{9x^{27}} \!+\! \frac{1721664}{x^{20}} 
  \!+\! \frac{22955520}{13 x^{13}} \!+\! \frac{1912960}{x^6} 
  \!-\! 1639680 \frac{x-1}{x^7-1} 
  \!+\! 234240 I(x), \label{eq:solS2}
  \\[0.2cm]%%%%%%%%%%%%%%%%%%%%%%%%%
  g_2(x) &= - \frac{94330880}{9 x^{27}} \!+\! \frac{655872}{x^{20}} 
  \!+\! \frac{13117440}{13 x^{13}} \!+\! \frac{2186240}{x^6} 
  \!+\! 1873920 I(x). \notag
\end{alignat}
Here $c_\text{s}$ is an integral constant.
The details of the derivation and an explicit form of $I(x)$ can be found in appendix \ref{app:solve}.
There other integral constants are set to be zero so that the geometry becomes asymptotically flat.
Notice that $c_\text{s}$ can be absorbed by the redefinition of $r_s$ up to the linear order of $\gamma$,
so physical quantities do not depend on $c_\text{s}$.

Let us investigate the thermodynamics of the quantum black string.
The event horizon is located at 
$r_\text{horizon}=r_\text{s} - \frac{\gamma}{7 r_\text{s}^5} f_1(1)$, 
and the temperature is evaluated as
\begin{alignat}{3}
  T_\text{s} &= \frac{1}{4\pi} G_1^{-\frac{1}{2}} \frac{d F_1}{dr} \Big|_{r_\text{horizon}} \notag
  \\
  &= \frac{7}{4\pi r_\text{s}} \Big\{ 1 + \gamma \Big( \frac{8}{7} f_1(1) + \frac{1}{7} f'_1(1) 
  - \frac{1}{2} g_1(1) \Big) \frac{1}{r_\text{s}^6} \Big\}
  \equiv \frac{7}{4 \pi} \bar{T}_\text{s}. \label{eq:T_s}
\end{alignat}
By solving this inversely, $r_\text{s}$ is expressed as
\begin{alignat}{3}
  r_\text{s} &= \frac{1}{\bar{T}_\text{s}} 
  \Big\{1 +  \gamma \Big( \frac{8}{7} f_1(1) + \frac{1}{7} f'_1(1) - \frac{1}{2} g_1(1) \Big) 
  \bar{T}_\text{s}^6 \Big\}. \label{eq:r_s}
\end{alignat}
From this relation, $r_s$ is replaced with the temperature when we calculate physical quantities
up to the linear order of $\gamma$.
For example, the location of the event horizon is given by
\begin{alignat}{3}
  r_\text{horizon} &= \frac{1}{\bar{T}_\text{s}} \Big\{1 
  + \gamma \Big( f_1(1) + \frac{1}{7} f'_1(1) - \frac{1}{2} g_1(1) \Big) \bar{T}_\text{s}^6 \Big\} \notag
  \\
  &= \frac{1}{\bar{T}_\text{s}} \Big\{1 
  - \gamma \Big( \frac{587024224}{117} + 117120\, I(1) \Big) \bar{T}_\text{s}^6 \Big\}. \label{eq:r_horizon2}
\end{alignat}
The location of the horizon moves inward and does not depends on $c_\text{s}$, just like the case of the
quantum black hole.

The ADM mass of the black string $M_\text{s}$ is evaluated as
\begin{alignat}{3}
  \frac{2\kappa_{10}^2}{8 V_{S^8}} M_\text{s} &= r_\text{s}^7 
  \Big( 1 + \gamma \frac{1639680 - c_\text{s}}{r_\text{s}^6} \Big) \notag
  \\
  &= \frac{1}{\bar{T}_\text{s}^7} 
  \Big\{ 1 + \gamma \Big( 1639680 + 8 f_1(1) + f'_1(1) - \frac{7}{2} g_1(1) - c_\text{s} \Big) 
  \bar{T}_\text{s}^6 \Big\} \notag
  \\
  &= \frac{1}{\bar{T}_\text{s}^7} \big( 1 - 4919040 \gamma \bar{T}_\text{s}^6 \big)
  \equiv \bar{M}_\text{s}. \label{eq:M_s}
\end{alignat}
Note that we used $2\kappa_{10}^2 = 2\kappa_{11}^2/(2\pi R_{11})$.
The mass formula itself is not modified, but the corrections in the first line enter through 
$x^{-7}$ terms in $f_1(x)$ and $g_2(x)$.
On the other hand, the entropy $S_\text{s}$ of the quantum black string receives 
the higher derivative corrections, and is calculated as
\begin{alignat}{3}
  \frac{2\kappa_{10}^2}{4\pi V_{S^8}} S_\text{s} &= r_\text{horizon}^8 G_2(1)^{1/2}
  \Big( 1 - 2 \gamma X_{0101} \Big|_{x=1} \Big) \notag
  \\
  &= \frac{1}{\bar{T}_\text{s}^8} \big( 1 - 2810880 \gamma \bar{T}_\text{s}^6 \big) \notag
  \\[0.2cm]
  &= \bar{M}_\text{s}^{8/7} \big( 1 + 2810880 \gamma \bar{M}_\text{s}^{-6/7} \big). \label{eq:S_s}
\end{alignat}
The entropy formula is affected by the higher derivative corrections.
Note that $r_\text{horizon}$, $M_\text{s}$ and $S_\text{s}$ do not depend on $c_s$,
and the first law of the thermodynamics $dM_\text{s} = T_\text{s} dS_\text{s}$ is satisfied
up to the linear of $\gamma$.

%%%%%%%%%%%%%%%%%%%%%%%%%%%%%%%%%%%%%%%%%%%%%%%%%%%%%%%%%%%%%%%%%%%%%%%%%%%%%%%%%%%%%%%%%%%%%%%%%%%%%
\subsection{Gregory-Laflamme Instability of the Quantum Black String}
\label{sec:ratio}

It is known that the black string is unstable against perturbation
when the size of the compactified direction is large compared to the size of the black hole.
In order to give thermodynamic argument on this instability, so called Gregory-Laflamme instability,
we compare $S_\text{h}$ with $S_\text{s}$ for the equal mass.

\begin{figure}[htbp]
\begin{center}
\begin{picture}(330,210)
\put(290,80){$\frac{S_\text{h}}{S_\text{s}}$}
\put(250,150){$L$}
\put(190,15){$g_s$}
\includegraphics[width=12cm]{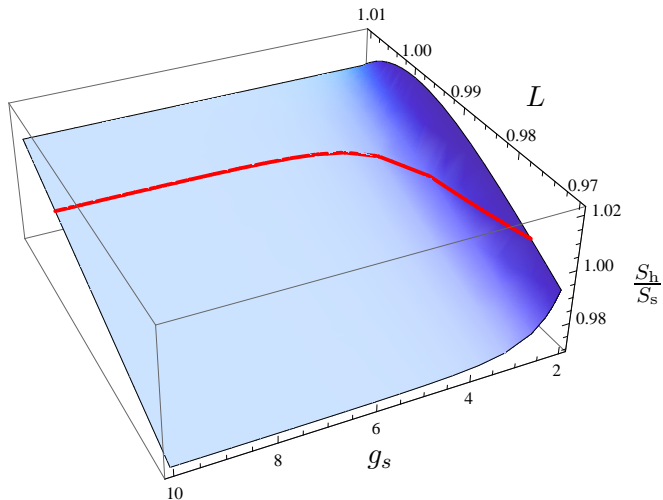}
\end{picture}
\caption{Plot of eq.~(\ref{eq:ratio}) which is valid when $L \leq 1$.
A red line correspond to $S_\text{h} = S_\text{s}$.
} \label{fig:ratio}
\end{center}
\end{figure}

In terms of a dimensionless parameter $M = \ell_s M_\text{h} = \ell_s M_\text{s}$, 
the entropy of the quantum black hole is given by
\begin{alignat}{3}
  S_\text{h} &= 8 \pi^2 \Big( \frac{g_s^3 M^9}{9^9 V_{S^9}} \Big)^{1/8} 
  \Big\{ 1 + \frac{10503}{2^{11} \pi^4} \frac{1}{g_s^4}
  \Big( \frac{9 V_{S^9} g_s^5}{M} \Big)^{3/4} \Big\}, \label{eq:S_h2}
\end{alignat}
and that of the quantum black string is given by
\begin{alignat}{3}
  S_\text{s} &= 8\pi^2 \Big( \frac{ g_s^2 M^8}{8^8 V_{S^8}} \Big)^{1/7}
  \Big\{ 1 + \frac{305}{2^7 \pi^4} \frac{1}{g_s^4}
  \Big( \frac{8 V_{S^8} g_s^5}{M} \Big)^{6/7} \Big\}. \label{eq:S_s2}
\end{alignat}
Here we used $2\kappa_{11}^2 = 2\kappa_{10}^2 (2\pi R_{11}) = (2\pi)^8 \ell_s^9 g_s^3$.
The instability of the black string is estimated by comparing these two entropies.
From eqs.~(\ref{eq:S_h2}) and (\ref{eq:S_s2}), up to the linear order of $\gamma$,
the ratio of the entropies is evaluated as
\begin{alignat}{3}
  \frac{S_\text{h}}{S_\text{s}} 
  &= L \bigg\{ 1 + \frac{10503}{2^{11} \pi^4} \Big( \frac{9^8 V_{S^9}}{8^8 V_{S^8}} \Big)^6 
  \frac{L^{42}}{g_s^4} - \frac{305}{2^7 \pi^4} \Big( \frac{9^9 V_{S^9}}{8^9 V_{S^8}} \Big)^6 
  \frac{L^{48}}{g_s^4} \bigg\} \notag
  \\
  &\sim L + 6.04 \frac{L^{43}}{g_s^4} - 5.69 \frac{L^{49}}{g_s^4}, \label{eq:ratio}
\end{alignat}
where we defined 
\begin{alignat}{3}
  L \equiv \frac{(8^8 V_{S^8})^{1/7}}{(9^9 V_{S^9})^{1/8}} \Big( \frac{g_s^5}{M} \Big)^{1/56}
  \sim 0.984 \Big( \frac{g_s^5}{M} \Big)^{1/56}. \label{eq:L}
\end{alignat}
In the classical supergravity limit, if we increase $L$ from zero, 
the Gregory-Laflamme transition from the black string to the black hole occurs at $L=1$.
By taking into account the quantum effect, the transition point also depends on the value of $g_s$ like
\begin{alignat}{3}
  L = 1 - \frac{0.350}{g_s^4}. \label{eq:Ltr1}
\end{alignat}
The plot of $S_\text{h}/S_\text{s}$ is drawn in fig.~\ref{fig:ratio}.

Finally let us clarify the validity of the approximation (\ref{eq:ratio}) qualitatively.
So far we have analyzed the black hole and black string solutions
by considering quantum corrections (\ref{eq:R4}) in 11 dimensions.
Since the 11th direction is compactified, the effective action corresponds to the type IIA superstring theory
and expanded by $g_s^2 e^{2\phi}$ and $\ell_s^2$.
Then we can examine the validity of eq.~(\ref{eq:ratio}) by estimating the other higher derivative terms
in the type IIA superstring theory, 
which can be done by evaluating the dilaton $\phi$ and the Riemann tensor $R_{abcd}$ in 10 dimensions
from eq.~(\ref{eq:11BS}). 
From the standard relation $ds_\text{s}^2 = e^{-\frac{2}{3}\phi} ds_{10}^2 + e^{\frac{4}{3}\phi} dz^2$, 
we obtain
\begin{alignat}{3}
  g_s e^\phi = g_s, \qquad \ell_s^2 R_{abcd} \sim \frac{\ell_s^2}{r_\text{s}^2} 
  \sim \ell_s^2 \bar{M}_\text{s}^{-2/7},
\end{alignat}
at the horizon.
Thus a generic term is estimated as
\begin{alignat}{3}
  (g_s^2 e^{2\phi})^n (\ell_s^2 R_{abcd})^m &\sim g_s^{2n} \ell_s^{2m} \bar{M}_\text{s}^{-2m/7}
\sim g_s^{2n-4m/7} M^{-2m/7} \sim g_s^{2n-2m} L^{16m}.
\end{alignat}
It is known that the leading tree, 1-loop and $n (\geq 2)$-loop corrections in the type IIA superstring theory 
become $(n,m)=(0,3), (1,3), (n,n+3)$, respectively\cite{Green:2006gt}.
So the estimations of leading tree, 1-loop and $n(\geq 2)$-loop are given by 
$g_s^{-6} L^{48}$, $g_s^{-4} L^{48}$ and $g_s^{-6} L^{16n+48}$, respectively.
In a similar way, from the dimensional analysis, the quantum corrections for the quantum black hole are estimated 
like $g_s^{2n} \ell_s^{2m} \bar{M}_\text{h}^{-m/4} \sim g_s^{2n-3m/4} M^{-m/4} \sim g_s^{2n-2m} L^{14m}$.

From the above discussions, we expect that the ratio of the entropies (\ref{eq:ratio}) is reliable when 
$L \leq 1$ and $1 \ll g_s^2$, if each term has a coefficient of order unity.
It is interesting to note that, around the transition point $L \sim 1$, the supergravity approximation is valid 
when $g_s$ goes to infinity, and quantum effects become quite important around $g_s \sim 1$.

%%%%%%%%%%%%%%%%%%%%%%%%%%%%%%%%%%%%%%%%%%%%%%%%%%%%%%%%%%%%%%%%%%%%%%%%%%%%%%%%%%%%%%%%%%%%%%%%%%%%%%%%%%%%%
%%%%%%%%%%%%%%%%%%%%%%%%%%%%%%%%%%%%%%%%%%%%%%%%%%%%%%%%%%%%%%%%%%%%%%%%%%%%%%%%%%%%%%%%%%%%%%%%%%%%%%%%%%%%%
\section{Boosted Quantum Black Hole and Black String in M-theory}\label{sec:boost}

%%%%%%%%%%%%%%%%%%%%%%%%%%%%%%%%%%%%%%%%%%%%%%%%%%%%%%%%%%%%%%%%%%%%%%%%%%%%%%%%%%%%%%%%%%%%%%%%%%%%%%%%%%%%%
\subsection{Boosted Quantum Black Hole}

From an observer at infinity, the quantum black hole is localized at the origin of 10 dimensional space. 
Now let us Lorentz boost the quantum black hole along the 11th direction.
Then the observer see that the quantum black hole carries a momentum along the 11th direction.

First the mass $M_\text{h}$ and the momentum $Q_\text{h}$ of the boosted quantum black hole are simply given by
\begin{alignat}{3}
  \frac{2\kappa_{11}^2}{9 V_{S^9}} M_\text{h} &= r_\text{h}^8 \cosh\eta
  \Big( 1 - \gamma \frac{c_\text{h}}{r_\text{h}^6} \Big) \equiv \bar{M}_\text{h}, \notag
  \\
  \frac{2\kappa_{11}^2}{9 V_{S^9}} Q_\text{h} &= r_\text{h}^8 \sinh\eta
  \Big( 1 - \gamma \frac{c_\text{h}}{r_\text{h}^6} \Big) \equiv \bar{Q}_\text{h}, \label{eq:M_bh}
\end{alignat}
where $\eta$ is a boost parameter\footnote{We assigned the same symbol $M_\text{h}$ as 
in the section \ref{sec:QBH}, since it would not cause any confusion.}.
By combining these two equations, the parameter of the quantum black hole is expressed as
\begin{alignat}{3}
  r_\text{h}^8 \Big( 1 - \gamma \frac{c_\text{h}}{r_\text{h}^6} \Big) &= \bar{M}_\text{h}
  \Big( 1 - \frac{\bar{Q}_\text{h}^2}{\bar{M}_\text{h}^2} \Big)^{1/2}. \label{eq:r_bh}
\end{alignat}

Next we examine the entropy of the boosted quantum black hole.
It is known that the volume of the event horizon is invariant under the Lorentz boost\cite{Horowitz:1997fr}.
The higher derivative term $X_{0101} = \frac{1}{4} N^{ab}N^{cd}X_{abcd}$ is invariant as well.
Therefore the entropy formula is entirely invariant under the Lorentz boost
and given by eq.~(\ref{eq:S_h}) by replacing $\bar{M}_\text{h}$ with the right hand side of eq.~(\ref{eq:r_bh}).
\begin{alignat}{3}
  \frac{2\kappa_{11}^2}{4\pi V_{S^9}} S_\text{h} &= \bar{M}_\text{h}^{9/8}
  \Big( 1 - \frac{\bar{Q}_\text{h}^2}{\bar{M}_\text{h}^2} \Big)^{9/16}
  \bigg\{ 1 + 6049728 \gamma \bar{M}_\text{h}^{-3/4} 
  \Big( 1 - \frac{\bar{Q}_\text{h}^2}{\bar{M}_\text{h}^2} \Big)^{-3/8} \bigg\}. \label{eq:S_bh}
\end{alignat}
Of course, eq.~(\ref{eq:S_h}) is recovered when $\bar{Q}_\text{h}=0$.

%%%%%%%%%%%%%%%%%%%%%%%%%%%%%%%%%%%%%%%%%%%%%%%%%%%%%%%%%%%%%%%%%%%%%%%%%%%%%%%%%%%%%%%%%%%%%%%%%%%%%%%%%%%%%%%%%%%%
\subsection{Boosted Quantum Black String}

Let us briefly review the Lorentz boost of the black string (\ref{eq:11BS}) along 11th direction. 
The boost is executed on $(t,z)$-plane and the metric becomes 
\begin{alignat}{3}
  ds_\text{s}^2 &= - F (\cosh \beta dt + \sinh \beta dz)^2 + (\sinh \beta dt + \cosh \beta dz)^2 
  + F^{-1} dr^2 + r^2 d\Omega_8^2 \notag
  \\
  &= - H^{-1} F dt^2 + H \Big( dz + (1 - H^{-1}) \frac{\cosh\beta}{\sinh \beta} dt \Big)^2 
  + F^{-1} dr^2 + r^2 d\Omega_8^2, \label{eq:11BS-b}
\end{alignat}
where $\beta$ is a boost parameter. Here $H$ is defined as
\begin{alignat}{3}
  H = 1 + \frac{r_\text{s}^7 \sinh^2 \beta}{r^7},
\end{alignat}
and $F$ is defined in eq.~(\ref{eq:11BS}).
This geometry corresponds to the nonextremal M-wave solution in 11 dimensional supergravity,
whose mass and momentum along the 11th direction are expressed by two parameters $r_s$ and $\beta$.
We also use $\alpha^7 \equiv 1/\sinh^2\beta$
for the boost parameter\footnote{These are related to $r_\pm$ as
$r_-^7 = r_\text{s}^7 \sinh^2\beta, \, r_+^7 = r_-^7 (1 + \alpha^7) = r_\text{s}^7 \cosh^2\beta$.}.
Note that the nonextremal M-wave solution is identified with the nonextremal black 0-brane solution
in 10 dimensions.

In the same way, it is possible to boost the quantum black string solution (\ref{eq:bsansatz}) 
along the 11th direction, and the metric becomes
\begin{alignat}{3}
  ds_\text{s}^2 &= - G_1^{-1} F_1 (\cosh \beta dt + \sinh \beta dz)^2 
  + G_2 (\sinh \beta dt + \cosh \beta dz)^2 + F_1^{-1} dr^2 + r^2 d\Omega_8^2 \notag
  \\
  &= - H_1^{-1} F_1 dt^2 + H_2 \Big( dz + \big( 1 - H_2^{-\frac{1}{2}} H_3^{-\frac{1}{2}} \big) 
  \frac{\cosh\beta}{\sinh \beta} dt \Big)^2 + F_1^{-1} dr^2 + r^2 d\Omega_8^2. \label{eq:met_bs}
\end{alignat}
Here $F_1$ is defined in eq.~(\ref{eq:bsansatz}), and
$H_i \,(i=1,2,3)$ are expressed in terms of $F_1$, $G_1$ and $G_2$ as
\begin{alignat}{3}
  H_1 = G_1 G_2^{-1} H_2, \qquad
  H_2 = G_2 + (G_2 - G_1^{-1} F_1) \sinh^2\beta, \qquad
  H_3 = G_2^{-2} H_2. \label{eq:Hs}
\end{alignat}
Now we choose $r_\text{s}$ and $\alpha$ as independent parameters.
Then the above functions can be expressed up to the linear order of $\gamma$ as
\begin{alignat}{3}
  H_1 &= 1 + \frac{1}{\alpha^7 x^7} + \frac{\gamma}{r_\text{s}^6 \alpha^7}
  \Big\{ - f_1(x) + ( 1 + \alpha^7 ) g_1(x) + \Big( 1 - \frac{1}{x^7} \Big) g_2(x) \Big\}, \notag
  \\
  H_2 &= 1 + \frac{1}{\alpha^7 x^7} + \frac{\gamma}{r_\text{s}^6 \alpha^7}
  \Big\{ - f_1(x) + \Big( 1 - \frac{1}{x^7} \Big) g_1(x) + (1 + \alpha^7) g_2(x) \Big\}, \label{eq:Hs2}
  \\
  H_3 &= 1 + \frac{1}{\alpha^7 x^7} + \frac{\gamma}{r_\text{s}^6 \alpha^7}
  \Big\{ - f_1(x) + \Big( 1 - \frac{1}{x^7} \Big)  g_1(x) + \Big( 1 - \alpha^7 - \frac{2}{x^7} \Big) g_2(x) 
  \Big\}, \notag
\end{alignat}
where $x = \frac{r}{r_\text{s}}$.
By inserting eq.~(\ref{eq:solS2}) and $c_\text{s} = 3747840$ into the above, we obtain eq.~(48) 
in ref. \cite{Hyakutake:2014maa}. 
In that paper $c_\text{s} = 3747840$ is required so as to be consistent with the near horizon limit
which will be explained in section \ref{sec:nearH}.
Thus the geometry (\ref{eq:met_bs}) is exactly the same as that of the quantum M-wave solution in 11 dimensions. 
The dimensional reduction of the metric corresponds to the quantum black 0-brane solution in 10 dimensions.

Let us examine the thermodynamics of the boosted quantum black string by choosing $c_\text{s} = 3747840$.
The event horizon is located at $r_\text{horizon}=r_\text{s} - \frac{\gamma}{7 r_\text{s}^5} f_1(1)$, 
and the temperature is evaluated as
\begin{alignat}{3}
  T_\text{s} &= \frac{1}{4\pi} H_1^{-1/2} \frac{d F_1}{dr} \Big|_{r_\text{horizon}} \notag
  \\
  &= \frac{7}{4\pi r_\text{s}} \frac{\alpha^{7/2}}{\sqrt{1+\alpha^7}} 
  \Big\{ 1 + \gamma \Big( \frac{8}{7} f_1(1) + \frac{1}{7} f'_1(1) 
  - \frac{1}{2} g_1(1) \Big) \frac{1}{r_\text{s}^6} \Big\} \notag
  \\
  &= \frac{7}{4\pi r_\text{s}} \frac{\alpha^{7/2}}{\sqrt{1+\alpha^7}} 
  \Big( 1 - \frac{2810880 \gamma}{7 r_\text{s}^6} \Big). \label{eq:T_bs}
\end{alignat}
Note that $\frac{\alpha^{7/2}}{\sqrt{1+\alpha^7}} = \frac{1}{\cosh \beta}$ comes from the time dilation of
the quantum black string measured by the boosted observer.

The ADM mass and the momentum of the boosted quantum black string are calculated as~\cite{Hyakutake:2014maa}
\begin{alignat}{3}
  \frac{2\kappa_{10}^2}{8 V_{S^8}} M_\text{s} &= r_\text{s}^7
  \Big( 1 + \frac{7}{8\alpha^7} - \frac{2108160 \gamma}{r_\text{s}^6} \Big) \equiv \bar{M}_\text{s}, \notag
  \\
  \frac{2\kappa_{10}^2}{8 V_{S^8}} Q_\text{s} &= \frac{7 r_\text{s}^7 \sqrt{1 + \alpha^7}}{8\alpha^7}
  \equiv \bar{Q}_\text{s}. \label{eq:M_bs}
\end{alignat}
The momentum does not receive any quantum correction and is equal to the charge of $N$ D0-branes.
By solving eqs.~(\ref{eq:M_bs}) inversely, up to the linear order of $\gamma$,
the parameters $\alpha^7$ and $r_\text{s}^7$ can be expressed as
\begin{alignat}{3}
  \alpha^7 &= \Big( \frac{7p}{8} \Big)^2 - 1, \qquad
  r_\text{s}^7 &= \bar{Q}_\text{s} \, p \Big( 1 - \frac{64}{49 p^2} \Big), \label{eq:alphars}
\end{alignat}
where
\begin{alignat}{3}
  &p = p_0 + 2108160 \gamma \, p_1, \notag
  \\[0.2cm]
  &p_0 = \frac{\bar{M}_\text{s}}{2 \bar{Q}_\text{s}} 
  \bigg( 1 + \sqrt{1 + \frac{32 \bar{Q}_\text{s}^2}{49 \bar{M}_\text{s}^2}} \bigg), \label{eq:p}
  \\
  &p_1 = \bar{Q}_\text{s}^{-6/7} \Big( 1 + \frac{8}{49 p_0^2} \Big)^{-1}
  \bigg\{ p_0 \Big( 1 - \frac{64}{49 p_0^2} \Big) \bigg\}^{1/7}. \notag
\end{alignat}
It is easy to see that $p_0 \to \infty$ and $Q_\text{s} p_0 \to M_\text{s}$, if we take $Q_\text{s} \to 0$.

The entropy of the boosted quantum black string is obtained by taking into account 
an expansion of the proper length along the 11th direction at the event horizon.
From eqs.~(\ref{eq:met_bs}) and (\ref{eq:Hs}), the expansion rate of the proper length 
at the event horizon is given by $\sqrt{H_2/G_2}|_{\text{horizon}} = \sqrt{1+\alpha^7}/\alpha^{7/2}$.
Thus, by multiplying this factor with eq.~(\ref{eq:S_s}), we obtain the entropy of the boosted quantum black 
string like
\begin{alignat}{3}
  \frac{2\kappa_{10}^2}{4\pi V_{S^8}} S_\text{s} &= r_\text{horizon}^8 \frac{\sqrt{1+\alpha^7}}{\alpha^{7/2}}
  G_2(1)^{1/2} \Big( 1 - 2 \gamma X_{0101} \Big|_{x=1} \Big) \notag
  \\
  &= r_\text{s}^8 \frac{\sqrt{1+\alpha^7}}{\alpha^{7/2}}
  \Big( 1 + \frac{2810880 \gamma}{7 r_\text{s}^6} \Big) \notag
  \\
  &= \big( \bar{Q}_\text{s} p_0 \big)^{8/7} \Big( 1 - \frac{64}{49 p_0^2} \Big)^{9/14} \bigg\{ 1 + 2810880 \gamma
  \big( \bar{Q}_\text{s} p_0 \big)^{-6/7} \Big( 1 - \frac{64}{49 p_0^2} \Big)^{-6/7} \bigg\}. \label{eq:S_bs}
\end{alignat}
In order to derive the second line, we used $c_\text{s} = 3747840$.
This satisfies the first law of the black hole thermodynamics up to the linear order of $\gamma$.

%%%%%%%%%%%%%%%%%%%%%%%%%%%%%%%%%%%%%%%%%%%%%%%%%%%%%%%%%%%%%%%%%%%%%%%%%%%%%%%%%%%%%%%%%%%%%%%%%%%%%%%%%%%%%%%%
\subsection{Gregory-Laflamme Instability of the Boosted Quantum Black String}

The Gregory-Laflamme instability of the boosted black string is well discussed 
in refs.~\cite{Horowitz:1997fr,Hovdebo:2006jy}.
In this subsection, we examine the Gregory-Laflamme instability of the boosted quantum black string.
In order to compare the entropy of the boosted black hole with that of the boosted quantum black string,
masses and charges should be equal respectively. 

\begin{figure}[htbp]
\begin{center}
\begin{picture}(280,200)
\put(250,70){$\frac{S_\text{h}}{S_\text{s}}$}
\put(210,143){$g_s$}
\put(145,1){$L$}
\includegraphics[width=9cm]{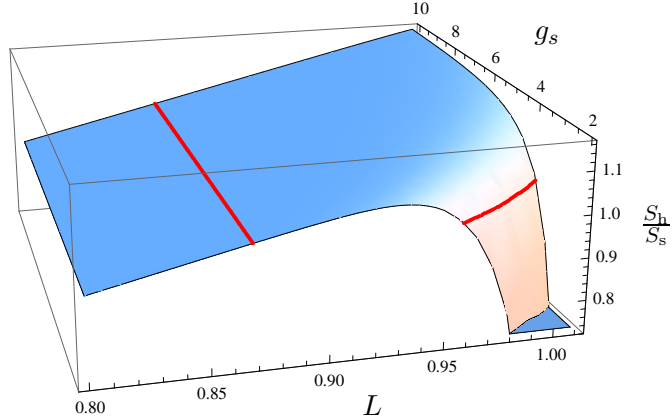}
\end{picture}
\caption{Plot of eq.~(\ref{eq:b-ratio}) with $\frac{Q}{M} = 0.9$. 
Red lines correspond to $S_\text{h} = S_\text{s}$.
} \label{fig:ratio2}
\end{center}
\end{figure}

Now we set $M = \ell_s M_\text{h} = \ell_s M_\text{s}$ and $Q = \ell_s Q_\text{h} = \ell_s Q_\text{s}$,
and use the relation $2\kappa_{11}^2 = 2\kappa_{10}^2 (2\pi R_{11}) = (2\pi)^8 \ell_s^9 g_s^3$.
Then the entropy of the boosted quantum black hole is expressed as
\begin{alignat}{3}
  S_\text{h} &= 8\pi^2 \Big( \frac{g_s^3 M^9}{9^9 V_{S^9}} \Big)^{1/8} \Big( 1 - \frac{Q^2}{M^2} \Big)^{9/16} 
  \bigg\{ 1 + \frac{10503}{2^{11} \pi^4} \frac{1}{g_s^4} \Big( \frac{9 V_{S^9} g_s^5}{M} \Big)^{3/4}
  \Big( 1 - \frac{Q^2}{M^2} \Big)^{-{3/8}} \bigg\}. \label{eq:S_bh2}
\end{alignat}
This is a generalization of eq.~(\ref{eq:S_h2}).
In a similar way, the entropy of the boosted quantum black string is given by
\begin{alignat}{3}
  S_\text{s} &= 8\pi^2 \Big( \frac{g_s^2 M^8}{8^8 V_{S^8}} \Big)^{1/7} \Big( \frac{Q p_0}{M} \Big)^{8/7}
  \Big( 1 - \frac{64}{49 p_0^2} \Big)^{9/14} \notag
  \\
  &\quad\, \times
  \bigg\{ 1 + \frac{305}{2^7 \pi^4} \frac{1}{g_s^4} \Big( \frac{8 V_{S^8} g_s^5}{M} \Big)^{6/7}
  \Big( \frac{Q p_0}{M} \Big)^{-6/7} \Big( 1 - \frac{64}{49 p_0^2} \Big)^{-6/7} \bigg\}. \label{eq:S_bs2}
\end{alignat}
Here $p_0$ is defined in eq. (\ref{eq:p}) and written in terms of $\frac{Q}{M}$.
This expression is a generalization of eq.~(\ref{eq:S_s2}), and
the ration of the entropies (\ref{eq:ratio}) is also generalized as
\begin{alignat}{3}
  \frac{S_\text{h}}{S_\text{s}} 
  &\sim L \Big( 1 - \frac{Q^2}{M^2} \Big)^{9/16} \Big( \frac{Q p_0}{M} \Big)^{-8/7} 
  \Big( 1 - \frac{64}{49 p_0^2} \Big)^{-9/14} \notag
  \\
  &\quad\, \times
  \bigg\{ 1 + 6.04 \frac{L^{42}}{g_s^4} \Big( 1 - \frac{Q^2}{M^2} \Big)^{-3/8} 
  - 5.69 \frac{L^{48}}{g_s^4} \Big( \frac{Q p_0}{M} \Big)^{-6/7} \Big( 1 - \frac{64}{49 p_0^2} \Big)^{-6/7} \bigg\},
  \label{eq:b-ratio}
\end{alignat}
where $L$ is defined by eq.~(\ref{eq:L}).
The ratio of the entropies are parametrized by $L$, $g_s$ and $\frac{Q}{M}$.
From this equation, we can estimate the instability of the boosted quantum black string.
The plot of eq.~(\ref{eq:b-ratio}) with $\frac{Q}{M} = 0.9$ is drawn in fig.~\ref{fig:ratio2}.
From the plot we see that, at transition point, the value of $L$ for the boosted quantum black string is smaller
than that of the quantum black string.
It is also interesting to note that there is another transition point for $g_s < 4$,
though the validity of it depends on the structure of the higher derivative terms as discussed below.

Finally let us examine the validity of eq.~(\ref{eq:b-ratio}) by estimating the other higher derivative terms.
In order to do this, we simply repeat the discussions in the section \ref{sec:ratio}.
From eq.~(\ref{eq:11BS-b}), values of the dilaton $\phi$ and the Riemann tensor $R_{abcd}$
in 10 dimensions are estimated as
\begin{alignat}{3}
  &g_s e^\phi = g_s \Big( 1 + \frac{1}{\alpha^7} \Big)^{3/4}
  = g_s \Big( 1 - \frac{64}{49 p_0^2} \Big)^{-3/4}, \notag
  \\
  &\ell_s^2 R_{abcd} \sim \frac{\ell_s^2}{r_\text{s}^2} w_{abcd}(p_0) \Big( 1 + \frac{1}{\alpha^7} \Big)^{-1/2}
  \sim \ell_s^2 w_{abcd}(p_0) (\bar{Q}_\text{s} p_0)^{-2/7} \Big( 1 - \frac{64}{49 p_0^2} \Big)^{3/14},
  \label{eq:gsR}
\end{alignat}
at the horizon.
Here explicit form of $w_{abcd}(p_0)$ is different for each component,
but becomes constant if we take $\frac{Q}{M} \to 1$.
From the above a generic term is estimated as
\begin{alignat}{3}
  (g_s^2 e^{2\phi})^n (\ell_s^2 R_{abcd})^m &\sim g_s^{2n} \ell_s^{2m} \bar{M}_\text{s}^{-2m/7}
  \Big( \frac{Q p_0}{M} \Big)^{-2m/7} (w_{abcd})^m
  \Big( 1 - \frac{64}{49 p_0^2} \Big)^{3(m-7n)/14} \notag
  \\
  &\sim g_s^{2n-2m} L^{16m} (w_{abcd})^m
  \Big( \frac{Q p_0}{M} \Big)^{-2m/7} \Big( 1 - \frac{64}{49 p_0^2} \Big)^{3(m-7n)/14} \notag
  \\
  &\sim g_s^{2n-2m} L^{16m},
\end{alignat}
where $L$ is defined by eq.~(\ref{eq:L}).
In the last line, we simply dropped $\frac{Q}{M}$ dependence, which deeply depends on the structure
of each higher derivative term.
Therefore we roughly estimate that the entropy of boosted quantum black string is valid when $L \leq 1$
and $1 \ll g_s^2$, if each higher derivative term has a coefficient of order unity. 
This will also be true for the boosted quantum black hole.

%%%%%%%%%%%%%%%%%%%%%%%%%%%%%%%%%%%%%%%%%%%%%%%%%%%%%%%%%%%%%%%%%%%%%%%%%%%%%%%%%%%%%%%%%%%%%%%
%%%%%%%%%%%%%%%%%%%%%%%%%%%%%%%%%%%%%%%%%%%%%%%%%%%%%%%%%%%%%%%%%%%%%%%%%%%%%%%%%%%%%%%%%%%%%%%
\section{Near Horizon Limit of the Boosted Quantum Black String and its Instability}
\label{sec:nearH}

The gauge/gravity correspondence is a powerful tool to investigate the nonperturbative aspects of 
the gauge theory\cite{Maldacena:1997re,Gubser:1998bc,Witten:1998qj}.
In this section we consider the near horizon limit of $N$ D0-branes, which corresponds to the infinitely
boosted ($\beta \to \infty$) quantum black string\cite{Itzhaki:1998dd}. 
The momentum of the boost corresponds to the charge of $N$ D0-branes, and is expressed as
\begin{alignat}{3}
  Q_\text{s} = \frac{N}{\ell_s g_s}. \label{eq:D0charge}
\end{alignat}
Since the boost parameter is given by $\alpha^7 = 1/\sinh^2\beta$,
the near horizon limit corresponds to $\alpha \to 0$.
While taking this limit, we should fix the 't Hooft coupling $\lambda$ and the energy scale $U_0$ of 
the gauge theory on $N$ D0-branes, which are written as
\begin{alignat}{3}
  \lambda = \frac{g_s N}{(2\pi)^2 \ell_s^3}, \qquad U_0 = \frac{r_\text{s}}{\ell_s^2}. \label{eq:lambda}
\end{alignat}
From eqs.~(\ref{eq:M_bs}) and (\ref{eq:D0charge}), $\alpha$ approaches to zero like
\begin{alignat}{3}
  \alpha^7 \;\to\; \frac{7 V_{S^8} \ell_s^4 U_0^7}{(2\pi)^9 \lambda}, \label{eq:alp-lim}
\end{alignat}
and $\ell_s$ also goes to zero in the limit.

Now we take the near horizon limit for the boosted quantum black string.
First of all, $Q_\text{s}$ is exactly given by eq.~(\ref{eq:D0charge}), 
so it is expressed in terms of $\ell_s$, $\lambda$ and $N$ as
\begin{alignat}{3}
  Q_\text{s} &= \frac{N^2}{(2\pi)^2 \ell_s^4 \lambda}. \label{eq:Q_bs-lim}
\end{alignat}
This approaches to the infinity by taking the near horizon limit.
Next, $M_\text{s}$ is given by eq.~(\ref{eq:M_bs}) and it behaves like
\begin{alignat}{3}
  M_\text{s} &= Q_\text{s} \frac{8\alpha^7}{7\sqrt{1+\alpha^7}}
  \Big( 1 + \frac{7}{8\alpha^7} - \frac{2108160 \gamma}{r_\text{s}^6} \Big) \notag
  \\
  &\to \frac{N^2}{(2\pi)^2 \ell_s^4 \lambda} \Big( 1 + \frac{3\ell_s^4 U_0^7}{35(2\pi)^5 \lambda}
  - \frac{61 \pi \ell_s^4 U_0 \lambda}{7 N^2} \Big). \label{eq:M_bs-lim}
\end{alignat}
By taking the near horizon limit, the leading term diverges like the charge.
However, the internal energy $E_\text{s} = M_\text{s} - Q_\text{s}$ becomes finite and is given by
\begin{alignat}{3}
  \tilde{E}_\text{s} &= \frac{3 N^2}{(2\pi)^7 35} \Big( \tilde{U}_0^7
  - \frac{9760 \pi^6 \tilde{U}_0}{3 N^2} \Big), \label{eq:E-lim}
\end{alignat}
where $\tilde{E}_\text{s} \equiv E_\text{s}/\lambda^{1/3}$ and $\tilde{U}_0 \equiv U_0/\lambda^{1/3}$.
The temperature (\ref{eq:T_bs}) approaches to
\begin{alignat}{3}
  \tilde{T}_\text{s} &= a_1 \tilde{U}_0^{5/2}
  \Big( 1 - \frac{2440 \pi^6}{7 N^2 \tilde{U}_0^6} \Big), \qquad
  a_1 \equiv \frac{7}{2^4 (15 \pi^7)^{1/2}}, \label{eq:T_bs-lim}
\end{alignat}
where $\tilde{T}_\text{s} \equiv T_\text{s}/\lambda^{1/3}$.
Inversely solving this, we obtain
\begin{alignat}{3}
  \tilde{U}_0 &= a_1^{-2/5} \tilde{T}_\text{s}^{2/5}
  \bigg( 1 + \frac{976 \pi^6 a_1^{12/5}}{7 N^2 \tilde{T}_\text{s}^{12/5}} \bigg). \label{eq:tildeU_0}
\end{alignat}
Thus physical quantities can be expressed in terms of $\tilde{T}_\text{s}$ and $N$.
Finally, the near horizon limit of the entropy (\ref{eq:S_bs}) is given by
\begin{alignat}{3}
  S_\text{s} &= \frac{N^2 \tilde{U}_0^{9/2}}{28 (15\pi^7)^{1/2}} 
  \bigg( 1 + \frac{2440 \pi^6}{7 N^2 \tilde{U}_0^6} \bigg) \notag
  \\
  &= \frac{4 N^2 \tilde{T}_\text{s}^{9/5}}{49 a_1^{4/5}} 
  \bigg( 1 + \frac{976 \pi^6 a_1^{12/5}}{N^2 \tilde{T}_\text{s}^{12/5}} \bigg) \notag
  \\
  &\sim 11.5 N^2 \tilde{T}_\text{s}^{9/5}
  \bigg( 1 + \frac{0.334}{N^2 \tilde{T}_\text{s}^{12/5}} \bigg). 
  \label{eq:S_bs2-lim}
\end{alignat}
This expression is the generalization of the discussion in ref.~\cite{Itzhaki:1998dd},
and first derived in ref.~\cite{Hyakutake:2013vwa} by evaluating in the background of the near horizon geometry. 

Let us examine the Gregory-Laflamme instability of the boosted quantum black string in the near horizon limit.
In order to do this, we need to compare the entropy (\ref{eq:S_bs2-lim}) 
with that of the boosted quantum black hole.
By taking the near horizon limit of eq.~(\ref{eq:S_bh2}), the entropy of the boosted quantum black hole 
is expressed as
\begin{alignat}{3}
  S_\text{h} &= \frac{N^{15/8} \tilde{U}_0^{63/16}}{3 \sqrt{2} (105)^{9/16} \pi^{47/16}} 
  \bigg( 1 - \frac{1830 \pi^6}{N^2 \tilde{U}_0^6} 
  + \frac{10503 (105)^{3/8} \pi^{21/8}}{2^{10} N^{5/4} \tilde{U}_0^{21/8}} \bigg) \notag
  \\
  &= \frac{N^{15/8} \tilde{T}_\text{s}^{63/40}}{3 \sqrt{2} (105)^{9/16} \pi^{47/16} a_1^{63/40}} 
  \bigg( 1 - \frac{1281 \pi^6 a_1^{12/5}}{N^2 \tilde{T}_\text{s}^{12/5}} 
  + \frac{10503 (105)^{3/8} \pi^{21/8} a_1^{21/20}}{2^{10} N^{5/4} \tilde{T}_\text{s}^{21/20}} \bigg) \notag
  \\[0.1cm]
  &\sim 10.2 N^{15/8} \tilde{T}_\text{s}^{63/40} 
  \bigg( 1 - \frac{0.438}{N^2 \tilde{T}_\text{s}^{12/5}} 
  + \frac{1.79}{N^{5/4} \tilde{T}_\text{s}^{21/20}} \bigg).
  \label{eq:S_bh2-lim}
\end{alignat}
It is worth noting that the leading part behaves like $N^{15/8}$ and the quantum effects consist of two terms.
It is challenging problem to explain these behaviors from the gauge theory on $N$ D0-branes.

The instability of the near horizon limit of the boosted quantum black string is investigated by
comparing eq.~(\ref{eq:S_bs2-lim}) with eq.~(\ref{eq:S_bh2-lim}).
Up to the next leading order, the ratio of the entropies is given by
\begin{alignat}{3}
  \frac{S_\text{h}}{S_\text{s}} &\sim L
  \bigg( 1 - \frac{2257 \pi^6 a_1^{12/5}}{N^2 \tilde{T}_\text{s}^{12/5}} 
  + \frac{10503 (105)^{3/8} \pi^{21/8} a_1^{21/20}}{2^{10} N^{5/4} \tilde{T}_\text{s}^{21/20}} \bigg) 
  \notag
  \\
  &\sim L
  \bigg( 1 - \frac{0.772}{N^2 \tilde{T}_\text{s}^{12/5}} 
  + \frac{1.79}{N^{5/4} \tilde{T}_\text{s}^{21/20}} \bigg), \notag
  \\
  &\sim L \Big( 1 - 2.93 \frac{L^{32/3}}{N^{2/3}} + 3.21 \frac{L^{14/3}}{N^{2/3}} \Big), \label{eq:n-ratio}
\end{alignat}
where
\begin{alignat}{3}
  L \equiv \frac{49}{12 \sqrt{2} (105)^{9/16} \pi^{47/16} a_1^{31/40} N^{1/8} \tilde{T}_\text{s}^{9/40}}
  \sim \frac{0.882}{N^{1/8} \tilde{T}_\text{s}^{9/40}}.
\end{alignat}
The transition of the boosted quantum black string in the near horizon limit occurs at
\begin{alignat}{3}
  L = 1 - \frac{0.277}{N^{2/3}}, \label{eq:Ltr2}
\end{alignat}
up to $\mathcal{O}(N^{-4/3})$.
This result is consistent with the discussion in ref.~\cite{Itzhaki:1998dd} in the classical limit.

\begin{figure}[htbp]
\begin{center}
\begin{picture}(330,210)
\put(290,100){$\frac{S_\text{h}}{S_\text{s}}$}
\put(220,175){$L$}
\put(200,15){$N$}
\includegraphics[width=11cm]{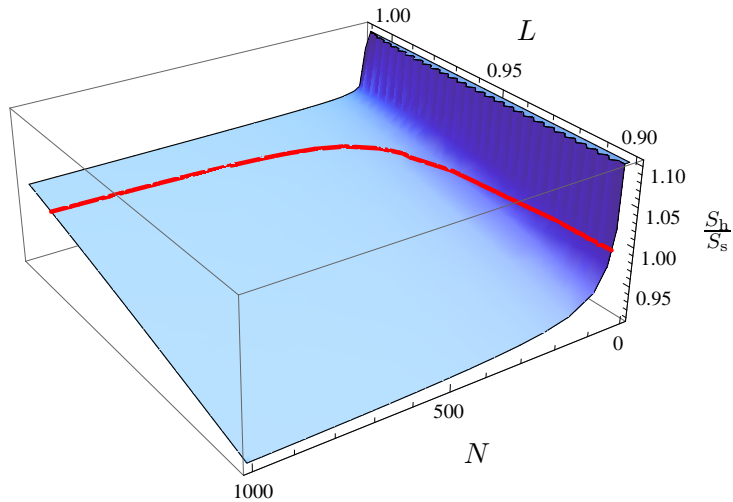}
\end{picture}
\caption{Plot of eq.~(\ref{eq:n-ratio}) which is valid when $L \leq 1$ and $L^{-56} \ll N$.
A red line correspond to $S_\text{h} = S_\text{s}$.
} \label{fig:ratio3}
\end{center}
\end{figure}

Let us examine the validity of eq.~(\ref{eq:n-ratio}) by estimating the other higher derivative terms.
The near horizon limit of eq.~(\ref{eq:gsR}) is given by
\begin{alignat}{3}
  &g_s e^\phi \sim \frac{\tilde{U}_0^{-21/4}}{N} \sim \frac{\tilde{T}_\text{s}^{-21/10}}{N}
  \sim N^{1/6} L^{28/3}, \notag
  \\
  &\ell_s^2 R_{abcd} \sim \tilde{U}_0^{3/2} \sim \tilde{T}_\text{s}^{3/5} \sim N^{-1/3} L^{-8/3},
\end{alignat}
and a generic term is estimated as
\begin{alignat}{3}
  (g_s^2 e^{2\phi})^n (\ell_s^2 R_{abcd})^m &\sim \frac{\tilde{T}_\text{s}^{(3m-21n)/5}}{N^{2n}}
  \sim N^{(n-m)/3} L^{8(7n-m)/3}.
\end{alignat}
The leading tree, 1-loop and $n (\geq 2)$-loop corrections in the type IIA superstring theory 
become $(n,m)=(0,3), (1,3), (n,n+3)$, respectively.
So the estimations of leading tree, 1-loop and $n(\geq 2)$-loop are given by 
$N^{-1}L^{-8}$, $N^{-2/3} L^{32/3}$ and $N^{-1} L^{16n-8}$, respectively.
Then the eq.~(\ref{eq:n-ratio}) is valid when
\begin{alignat}{3}
  L \leq 1, \qquad L^{-56} \ll N,
\end{alignat}
if each higher derivative term has a coefficient of order unity.
The plot of eq.~(\ref{eq:n-ratio}) is drawn in fig. \ref{fig:ratio3}.

%%%%%%%%%%%%%%%%%%%%%%%%%%%%%%%%%%%%%%%%%%%%%%%%%%%%%%%%%%%%%%%%%%%%%%%%%%%%%%%%%%%%%%%%%%%%%%%
%%%%%%%%%%%%%%%%%%%%%%%%%%%%%%%%%%%%%%%%%%%%%%%%%%%%%%%%%%%%%%%%%%%%%%%%%%%%%%%%%%%%%%%%%%%%%%%
\section{Conclusion and Discussion}\label{sec:conc}

In this paper we explored the quantum nature of the black hole and black string in 11 dimensions
by taking into account the higher derivative $R^4$ corrections.
Especially we clarified the transition from the quantum black string to the quantum black hole
from entropic arguments.

First we constructed the solutions of quantum black hole and black string up to the linear order 
of $\ell_p^6$. These are asymptotically flat, but the behaviors near the event horizons are
quite different from the classical ones.
We also investigated the thermodynamics of these quantum solutions, which satisfy the first law
of the thermodynamics.
The entropies of both quantum black hole and black string increase because of the quantum corrections.
By comparing these two, we discussed the quantum nature of the Gregory-Laflamme instability.
When the string coupling constant $g_s$ is quite large, it is reasonable to trust the classical analyses
and from eq.~(\ref{eq:Ltr1}) the transition occurs around
\begin{alignat}{3}
  M \sim 0.412 \, g_s^5 + 8.09 \, g_s. \label{eq:Mtrn}
\end{alignat}
On the other hand, when $g_s \sim 1$, we should take into account other higher derivative corrections
more seriously.
Notice also that we neglected the effect of the circle compactification on the black hole
to derive the above transition point.
The effect of the compactification on the black hole is investigated in ref.~\cite{Myers:1986rx},
and corrections to the Gregory-Laflamme instability are explored analytically
when the mass of the black hole is small in refs.~\cite{Kol:2003if,Harmark:2003yz}.
By consulting the result in ref.~\cite{Harmark:2003yz}, we see that the entropy (\ref{eq:S_h2}) is modified into
\begin{alignat}{3}
  S_\text{h} &= 8 \pi^2 \Big( \frac{g_s^3 M^9}{9^9 V_{S^9}} \Big)^{1/8} 
  \Big\{ 1 + \frac{10503}{2^{11} \pi^4} \Big( \frac{9^8 V_{S^9}}{8^8 V_{S^8}} \Big)^6 \frac{L^{42}}{g_s^4} 
  + \frac{\zeta(8)}{16 V_{S^9}} \frac{(8^8 V_{S^8})^8}{(9^9 V_{S^9})^7} \frac{1}{L^{56}} \Big\} \notag
  \\
  &\sim 8 \pi^2 \Big( \frac{g_s^3 M^9}{9^9 V_{S^9}} \Big)^{1/8} 
  \Big( 1 + 6.04 \frac{L^{42}}{g_s^4} + 0.00101 \frac{1}{L^{56}} \Big).
\end{alignat}
The third term in the parenthesis corresponds to the effect of the circle compactification.
Therefore when the parameters are in the region of $L \sim 1$ and $1 < g_s < 5$, 
the effect of the compactification is negligible compared to the quantum correction.
And when the parameters are in the region of $L \sim 1$ and $5 \leq g_s$, both effects are negligible
compared to the leading part. When the mass of the black hole is not small, we have to employ 
numerical calculation\cite{Sorkin:2003ka}.

Second we boosted the quantum black hole and black string solutions, and
showed that the latter corresponds to the nonextremal quantum black 0-brane solution.
Both entropies depend on the boost parameter in a complicated way, and because of this, the transition occurs
for larger $M$ than eq.~(\ref{eq:Mtrn}).
It is interesting to note that there appear another transition point around $g_s \sim 4$ for $\frac{Q}{M}=0.9$.

Finally we consider the near horizon limit of the boosted quantum black string.
In this limit, the boost parameter goes to the infinity, and physical quantities are
expressed in terms of the temperature. From eq.~(\ref{eq:Ltr2}) the transition occurs around
\begin{alignat}{3}
  T_\text{s} \sim \frac{0.574}{N^{5/9}} + \frac{0.707}{N^{11/9}}.
\end{alignat}
This shows that quantum effects become important when the number of D0-branes $N$ becomes small.

As a future work it is interesting to understand the Gregory-Laflamme instability in terms of the 
dual gauge theory. In fact numerical study of the thermal D0-branes system has been 
investigated considerably in refs.~\cite{KLL}-\cite{Kadoh:2015mka}, and especially 
the corresponding instability is numerically discussed in ref.~\cite{Hanada:2013rga}.
It is also of great interest to understand the Gregory-Laflamme instability of the black string 
from the gauge theory side which is not stretching to the 11th direction\cite{Hyakutake:2001kn,Azuma:2014cfa}.
In this paper we focused on $g_s$ correction in the type IIA superstring theory,
but it seems to be possible to examine $\alpha'$ correction as well.
Finally the confirmations of the relation between thermodynamic and perturbative instabilities are 
important directions\cite{Gregory:1993vy,Gubser:2000mm,Figueras:2011he,Hollands:2012sf}.

%%%%%%%%%%%%%%%%%%%%%%%%%%%%%%%%%%%%%%%%%%%%%%%%%%%%%%%%%%%%%%%%%%%%%%%%%%%%%%%%%%%%%%%%%%%%%%%

\section*{Acknowledgement}

The author would like to thank Takanori Fujiwara, Yoshinobu Habara, Yosuke Imamura, Katsushi Ito, 
Tetsuji Kimura, Yusuke Kimura, Tatsuma Nishioka, Makoto Sakaguchi,
Hidehiko Shimada, Fumihiko Sugino and Shinya Tomizawa.
This work was partially supported by the Ministry of Education, Science, 
Sports and Culture, Grant-in-Aid for Young Scientists (B) 24740140, 2012.

\appendix

%%%%%%%%%%%%%%%%%%%%%%%%%%%%%%%%%%%%%%%%%%%%%%%%%%%%%%%%%%%%%%%%%%%%%%%%%%%%%%%%%%%%%%%%%%%%%%%
%%%%%%%%%%%%%%%%%%%%%%%%%%%%%%%%%%%%%%%%%%%%%%%%%%%%%%%%%%%%%%%%%%%%%%%%%%%%%%%%%%%%%%%%%%%%%%%
\section{Calculations on Quantum Black Hole}\label{sec:appA}

%%%%%%%%%%%%%%%%%%%%%%%%%%%%%%%%%%%%%%%%%%%%%%%%%%%%%%%%%%%%%%%%%%%%%%%%%%%%%%%%%%%%%%%%%%%%%%%
\subsection{Explicit Values of Tensors}

First we choose the vielbein of the quantum black hole as follows.
\begin{alignat}{3}
  e^0 &= r_\text{h} B_1^{-1/2} A_1^{1/2} d\tau, \quad
  e^1 = r_\text{h} A_1^{-1/2} dx, \quad 
  e^2 = r_\text{h} x d\theta_1, \notag
  \\
  e^3 &= r_\text{h} x \cos\theta_1 d\theta_2, \quad
  e^4 = r_\text{h} x \cos\theta_1 \cos\theta_2 d\theta_3, \quad
  e^5 = r_\text{h} x \cos\theta_1 \cos\theta_2 \sin\theta_3 d\theta_4, \notag
  \\
  e^6 &= r_\text{h} x \cos\theta_1 \sin\theta_2 d\theta_5, \quad
  e^7 = r_\text{h} x \cos\theta_1 \sin\theta_2 \sin\theta_5 d\theta_6, \quad
  e^8 = r_\text{h} x \sin\theta_1 d\theta_7,
  \\
  e^9 &= r_\text{h} x \sin\theta_1 \sin\theta_7 d\theta_8,\quad
  e^{10} = r_\text{h} x \sin\theta_1 \cos\theta_7 d\theta_9, \notag
\end{alignat}
where $A_1(x) = 1 - \frac{1}{x^8} + \frac{\gamma}{r_\text{h}^6} a_1 (x)$ and
$B_1(x) = 1 + \frac{\gamma}{r_\text{h}^6} b_1 (x)$.
Then, up to the linear order of $\gamma$, nonzero components of the Riemann tensor, Ricci tensor and
scalar curvature are calculated as
\begin{alignat}{3}
  &R_{0101} = - \frac{36}{r_\text{h}^2 x^{10}} + \gamma
  \frac{x^9 a_1'' + x (1-x^8) b_1'' - 12 b_1'}{2 r_\text{h}^8x^9}, \notag
  \\
  &R_{0\hat{i}0\hat{i}} = \frac{4}{r_\text{h}^2 x^{10}} + \gamma \frac{x^8 a_1' 
  + (1-x^8) b_1'}{2 r_\text{h}^8 x^9}, \notag
  \\
  &R_{1\hat{i}1\hat{i}} = - \frac{4}{r_\text{h}^2 x^{10}} - \gamma \frac{a_1'}{2 r_\text{h}^8 x}, \quad
  R_{\hat{i}\hat{j}\hat{i}\hat{j}} = \frac{1}{r_\text{h}^2 x^{10}} - \gamma \frac{a_1}{r_\text{h}^8 x^2}, \notag
  \\[0.2cm]
  &R_{00} = \gamma \frac{x^9 a_1'' + 9 x^8 a_1' + x (1-x^8) b_1'' - 3 (1 + 3 x^8) b_1'}{2 r_\text{h}^8 x^9},
  \\
  &R_{11} = \gamma \frac{- x^9 a_1'' - 9 x^8 a_1' - x (1-x^8) b_1'' + 12 b_1'}{2 r_\text{h}^8 x^9}, \notag
  \\
  &R_{\hat{i}\hat{i}} = \gamma \frac{ - 2 x^8 a_1' -16 x^7 a_1 - (1-x^8) b_1'}{2 r_\text{h}^8 x^9}, \notag
  \\[0.2cm]
  &R = \gamma \frac{ - x^9 a_1'' - 18 x^8 a_1' - 72 x^7 a_1 - x (1-x^8) b_1'' + 3 (1 + 3 x^8) b_1'}
  {r_\text{h}^8 x^9}, \notag
\end{alignat}
where $\hat{i},\hat{j}=2,\cdots,10$ and $\hat{i} \neq \hat{j}$.

Next we evaluate higher derivative terms up to $\mathcal{O}(\gamma)$.
Nonzero components of $X_{abcd}$, $RX_{ij} \equiv R_{abci}X^{abc}{}_j$ and 
$DDX_{ij} \equiv D_{(a} D_{b)} X^a{}_{ij}{}^b$ are evaluated as
\begin{alignat}{3}
  &X_{0101} = - \frac{44070912}{r_\text{h}^6 x^{30}}, \quad
  X_{0\hat{i}0\hat{i}} = - X_{1\hat{i}1\hat{i}} = - \frac{2844672}{r_\text{h}^6 x^{30}}, \quad
  X_{\hat{i}\hat{j}\hat{i}\hat{j}} = \frac{1949952}{r_\text{h}^6 x^{30}}, \notag
  \\
  &RX_{00} = - RX_{11} = - \frac{2968289280}{r_\text{h}^8 x^{40}}, \quad
  RX_{\hat{i}\hat{i}} = - \frac{14315520}{r_\text{h}^8 x^{40}}, \notag
  \\
  &DDX_{00} = - \frac{1902182400 (13 - 11 x^8)}{r_\text{h}^8 x^{40}}, \quad
  DDX_{11} = \frac{900 (3445248 + 781824 x^8)}{r_\text{h}^8 x^{40}}, 
  \\
  &DDX_{\hat{i}\hat{i}} = \frac{78182400 (31 - 23 x^8)}{r_\text{h}^8 x^{40}}, \quad
  t_8 t_8 R^4 - \frac{1}{4!} \epsilon_{11} \epsilon_{11} R^4 = \frac{1451934720}{r_\text{h}^8 x^{40}}, \notag
\end{alignat}
where $\hat{i},\hat{j}=2,\cdots,10$ and $\hat{i} \neq \hat{j}$.

%%%%%%%%%%%%%%%%%%%%%%%%%%%%%%%%%%%%%%%%%%%%%%%%%%%%%%%%%%%%%%%%%%%%%%%%%%%%%%%%%%%%%%%%%%%%%%%%%%%%%%%%
\subsection{Plots of $A_1(x)$}\label{sec:A1}

\begin{figure}[htbp]
\begin{center}
\begin{picture}(300,200)
\put(290,93){$x$}
\put(7,170){$A_1(x)$}
\put(60,160){$\tilde{\gamma}=10^{-8}$}
\put(96,110){$\tilde{\gamma}=10^{-9}$}
\includegraphics[width=10cm]{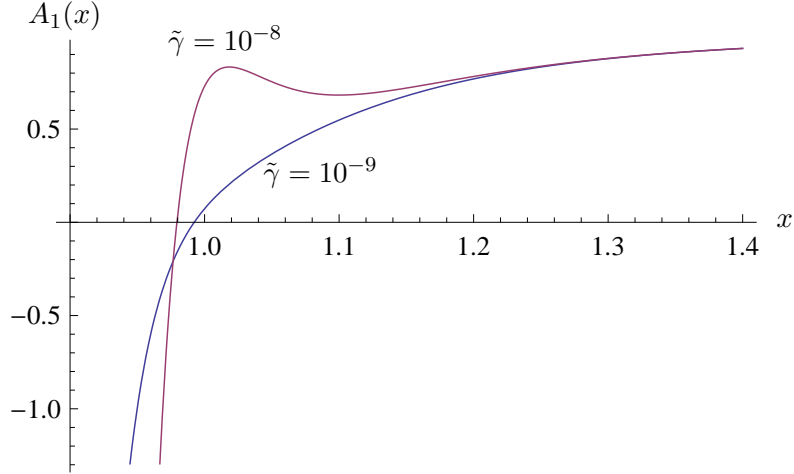}
\end{picture}
\caption{Plots of $A_1(x)$ with $\tilde{\gamma} = 10^{-9}$ and $10^{-8}$.
} \label{fig:A_1}
\end{center}
\end{figure}

Let us examine the properties of $A_1(x)$ given in section \ref{sec:QBH}.
For simplicity we just set $c_\text{h} = 0$ below, so $A_1(x)$ is given by
\begin{alignat}{3}
  A_1(x) = 1 - \frac{1}{x^8} + \tilde{\gamma} 
  \Big( \frac{422707200}{x^{30}} - \frac{349736448}{x^{38}} \Big), \qquad
  \tilde{\gamma} = \frac{\gamma}{r_\text{h}^6}.
\end{alignat}
The derivative of $A_1(x)$ is calculated as
\begin{alignat}{3}
  A_1'(x) = \frac{8 \tilde{\gamma}}{x^{39}} \big( \tilde{\gamma}^{-1} x^{30} - 1585152000 x^8 + 1661248128 \big).
\end{alignat}
Then $A_1'(x) = 0$ has one or two solutions when the minimum of the function in the parentheses becomes 
zero or negative, respectively. The function in the parentheses becomes minimum when
$x^{22} = 422707200 \tilde{\gamma}$, and the minimum takes negative value when
\begin{alignat}{3}
  - 1162444800 (422707200 \tilde{\gamma})^{4/11} + 1661248128 < 0 
  \;\;\Leftrightarrow\;\; 6.32 \times 10^{-9} < \tilde{\gamma}.
\end{alignat}
Plots of $A_1(x)$ with $\tilde{\gamma} = 10^{-9}$ and $10^{-8}$ are shown in fig.~\ref{fig:A_1}.
In both cases, locations of the event horizons are shifted inward compared with the classical case.
Especially the behavior of $A_1(x)$ with $\tilde{\gamma}=10^{-8}$ is quite different around the event horizon,
so a test particle feels a repulsive force.

%%%%%%%%%%%%%%%%%%%%%%%%%%%%%%%%%%%%%%%%%%%%%%%%%%%%%%%%%%%%%%%%%%%%%%%%%%%%%%%%%%%%%%%%%%%%%%%
%%%%%%%%%%%%%%%%%%%%%%%%%%%%%%%%%%%%%%%%%%%%%%%%%%%%%%%%%%%%%%%%%%%%%%%%%%%%%%%%%%%%%%%%%%%%%%%
\section{Calculations on Quantum Black String}\label{sec:appB}

%%%%%%%%%%%%%%%%%%%%%%%%%%%%%%%%%%%%%%%%%%%%%%%%%%%%%%%%%%%%%%%%%%%%%%%%%%%%%%%%%%%%%%%%%%%%%%%
\subsection{Explicit Values of Tensors}

First we choose the vielbein of the quantum black hole as follows.
\begin{alignat}{3}
  e^0 &= r_\text{s} G_1^{-1/2} F_1^{1/2} d\tau, \quad
  e^1 = r_\text{s} F_1^{-1/2} dx, \quad 
  e^2 = r_\text{s} x d\theta_1, \notag
  \\
  e^3 &= r_\text{s} x \cos\theta_1 d\theta_2, \quad
  e^4 = r_\text{s} x \cos\theta_1 \cos\theta_2 d\theta_3, \quad
  e^5 = r_\text{s} x \cos\theta_1 \cos\theta_2 \sin\theta_3 d\theta_4, \notag
  \\
  e^6 &= r_\text{s} x \cos\theta_1 \sin\theta_2 d\theta_5, \quad
  e^7 = r_\text{s} x \cos\theta_1 \sin\theta_2 \sin\theta_5 d\theta_6, \quad
  e^8 = r_\text{s} x \sin\theta_1 d\theta_7,
  \\
  e^9 &= r_\text{s} x \sin\theta_1 \sin\theta_7 d\theta_8,\quad
  e^{\natural} = r_\text{s} G_2^{1/2} dy, \notag
\end{alignat}
where $F_1(x) = 1 - \frac{1}{x^7} + \frac{\gamma}{r_\text{s}^6} f_1 (x)$ and
$G_i(x) = 1 + \frac{\gamma}{r_\text{s}^6} g_i (x)$.
Then, up to the linear order of $\gamma$, nonzero components of the Riemann tensor, Ricci tensor and
scalar curvature are calculated as
\begin{alignat}{3}
  &R_{0101} = - \frac{28}{r_\text{s}^2 x^9} + \gamma \frac{ 2 x^8 f_1'' 
  + 2 x (1-x^7) g_1'' - 21 g_1' }{4 r_\text{s}^8 x^8}, \quad
  R_{0\hat{i}0\hat{i}} = \frac{7}{2 r_\text{s}^2 x^9} + \gamma \frac{ x^7 f_1' 
  + (1-x^7) g_1'}{2 r_\text{s}^8 x^8}, \notag
  \\
  &R_{0\natural 0\natural} = \gamma \frac{7 g_2'}{4 r_\text{s}^8 x^8}, \quad
  R_{1\hat{i}1\hat{i}} = - \frac{7}{2 r_\text{s}^2 x^9} - \gamma \frac{f_1'}{2 r_\text{s}^8 x}, \quad
  R_{1\natural 1\natural} = \gamma \frac{ 2 x (1-x^7) g_2'' - 7 g_2'}{4 r_\text{s}^8 x^8}, \notag
  \\
  &R_{\hat{i}\hat{j}\hat{i}\hat{j}} = \frac{1}{r_\text{s}^2 x^9} - \gamma \frac{f_1}{r_\text{s}^8 x^2}, \quad
  R_{\hat{i}\natural \hat{i}\natural} = \gamma \frac{(1-x^7) g_2'}{2 r_\text{s}^8 x^8},
  \\[0.2cm]
  &R_{00} = \gamma \frac{ 2 x^8 f_1'' + 16 x^7 f_1' + 2 x (1-x^7) g_1'' 
  - (5 + 16 x^7) g_1' + 7 g_2'}{4 r_\text{s}^8 x^8}, \notag
  \\
  &R_{11} = \gamma \frac{ - 2 x^8 f_1'' -16 x^7 f_1' - 2 x (1-x^7) g_1'' + 21 g_1' 
  + 2 x (1-x^7) g_2'' - 7 g_2'}{4 r_\text{s}^8 x^8}, \notag
  \\
  &R_{\hat{i}\hat{i}} = \gamma \frac{ - 2 x^7 f_1' - 14 x^6 f_1 - (1-x^7) g_1' + (1-x^7) g_2'}{2
   r_\text{s}^8 x^8}, \quad
  R_{\natural\natural} = \gamma \frac{ x (1-x^7) g_2'' + (1 - 8 x^7) g_2'}{2 r_\text{s}^8 x^8}, \notag
  \\[0.2cm]
  &R = \gamma \frac{ - 2 x^8 f_1'' - 32 x^7 f_1' - 112 x^6 f_1 - 2 x (1-x^7) g_1''
  + (5 + 16 x^7) g_1' + 2 x (1-x^7) g_2'' + 2 (1 - 8 x^7) g_2'}{2 r_\text{s}^8 x^8}, \notag
\end{alignat}
where $\hat{i},\hat{j}=2,\cdots,9$ and $\hat{i} \neq \hat{j}$.

Next we evaluate higher derivative terms up to $\mathcal{O}(\gamma)$.
Nonzero components of $X_{abcd}$, $RX_{ij} \equiv R_{abci}X^{abc}{}_j$ and 
$DDX_{ij} \equiv D_{(a} D_{b)} X^a{}_{ij}{}^b$ are evaluated as
\begin{alignat}{3}
  &X_{0101} = - \frac{20321280}{r_\text{s}^6 x^{27}}, \quad
  X_{0\hat{i}0\hat{i}} = - X_{1\hat{i}1\hat{i}} = -\frac{1270080}{r_\text{s}^6 x^{27}}, \quad
  X_{\hat{i}\hat{j}\hat{i}\hat{j}} = \frac{1192320}{r_\text{s}^6 x^{27}}, \notag
  \\
  &RX_{00} = - RX_{11} = - \frac{1066867200}{r_\text{s}^8 x^{36}}, \quad
  RX_{\hat{i}\hat{i}} = - \frac{1088640}{r_\text{s}^8 x^{36}}, \notag
  \\
  &DDX_{00} = \frac{198132480 (- 47 + 40 x^7)}{r_\text{s}^8 x^{36}}, \quad
  DDX_{11} = \frac{1701 (1313280 + 317440 x^7)}{2 r_\text{s}^8 x^{36}}, 
  \\
  &DDX_{\hat{i}\hat{i}} = \frac{236234880 (4 - 3 x^7)}{r_\text{s}^8 x^{36}}, \quad
  t_8 t_8 R^4 - \frac{1}{4!} \epsilon_{11} \epsilon_{11} R^4 = \frac{531256320}{r_\text{s}^8 x^{36}}, \notag
\end{alignat}
where $\hat{i},\hat{j}=2,\cdots,9$ and $\hat{i} \neq \hat{j}$.

%%%%%%%%%%%%%%%%%%%%%%%%%%%%%%%%%%%%%%%%%%%%%%%%%%%%%%%%%%%%%%%%%%%%%%%%%%%%%%%%%%%%%%%%%%%%%%%%%%%%%%%%%%%%%
\subsection{Solution of eq.~(\ref{eq:EOM-s})}\label{app:solve}

Let us solve the equations of motion (\ref{eq:EOM-s}) for the quantum black string.
First we consider the following combinations.
\begin{alignat}{3}
  &\frac{E_1 + E_2}{16 x^{28}(1 - x^7)} = g_1' + \frac{1}{8} x g_2'' + \frac{4097640960}{x^{28}} = 0, \notag
  \\
  &\frac{E_1 + E_2 + 8(E_3 - E_4)}{2x^{28}} = - 16 x^7 f_1' - 112 x^6 f_1 
  - 7 x (1 - x^7) g_2'' + 56 x^7 g_2' \notag
  \\
  &\qquad\qquad\qquad\qquad\qquad\quad
  + \frac{2525644800}{x^{28}} - \frac{10102579200}{x^{21}} = 0.
\end{alignat}
These are solved as
\begin{alignat}{3}
  g_1 &= c_1 - \frac{1}{8} x g_2' + \frac{1}{8} g_2 + \frac{151764480}{x^{27}}, \notag
  \\
  f_1 &= \frac{c_2}{x^7} - \frac{7(1 - x^7)}{16 x^6}  g_2' + \frac{7}{16 x^7} g_2
  - \frac{5846400}{x^{34}} + \frac{31570560}{x^{27}}. \label{eq:solg1f1}
\end{alignat}
By inserting these solutions into $E_1=0$, we obtain
\begin{alignat}{3}
  \frac{E_1}{9 x^{28}} &= x (1 - x^7) g_2'' + (1 - 8 x^7) g_2' 
  + \frac{7640801280}{x^{28}} - \frac{5922201600}{x^{21}} \notag
  \\
  &= \Big\{ x (1 - x^7) g_2' - \frac{282992640}{x^{27}} + \frac{296110080}{x^{20}} \Big\}' = 0.
\end{alignat}
From this $g_2(x)$ is solved as
\begin{alignat}{3}
  g_2(x) &= c_3 + c_4 \log \frac{x^7}{x^7-1} \!-\! \frac{94330880}{9 x^{27}} \!+\! \frac{655872}{x^{20}} 
  \!+\! \frac{13117440}{13 x^{13}} \!+\! \frac{2186240}{x^6} \!+\! 1873920 I(x),
\end{alignat}
where 
\begin{alignat}{3}
  &I (x) = \log \frac{x^7(x-1)}{x^7-1} 
  - \sum_{n=1,3,5} \cos \frac{n\pi}{7} \log \Big( x^2 + 2 x \cos \frac{n\pi}{7} + 1 \Big) \notag
  \\[-0.1cm]
  &\quad\,
  - 2 \sum_{n=1,3,5} \sin \frac{n\pi}{7} \bigg\{
  \tan^{-1} \bigg( \frac{x + \cos \tfrac{n\pi}{7} }
  {\sin \tfrac{n\pi}{7}} \bigg) - \frac{\pi}{2} \bigg\}, \label{eq:I}
  \\
  &I'(x) = \frac{7(x-1)}{x(x^7-1)}. \notag
\end{alignat}
$c_3$ and $c_4$ are integral constants but should be zero 
so that the solution becomes asymptotically flat and $g_2(1)$ is finite.
$g_1(x)$ and $f_1(x)$ are determined by using eq.~(\ref{eq:solg1f1}).
$c_1$ should be zero because of the asymptotic flatness but $c_2$ remains as a constant parameter.
We have solved three out of four equations in (\ref{eq:EOM-s}), but the remaining equation 
is automatically satisfied.

%%%%%%%%%%%%%%%%%%%%%%%%%%%%%%%%%%%%%%%%%%%%%%%%%%%%%%%%%%%%%%%%%%%%%%%%%%%%%%%%%%%%%%%%%%%%%%%%%%%%%%%%%
\subsection{Plots of $F_1(x)$}

\begin{figure}[htbp]
\begin{center}
\begin{picture}(300,200)
\put(290,117){$x$}
\put(10,182){$F_1(x)$}
\put(60,175){$\tilde{\gamma}=10^{-7}$}
\put(75,131){$\tilde{\gamma}=10^{-8}$}
\includegraphics[width=10cm]{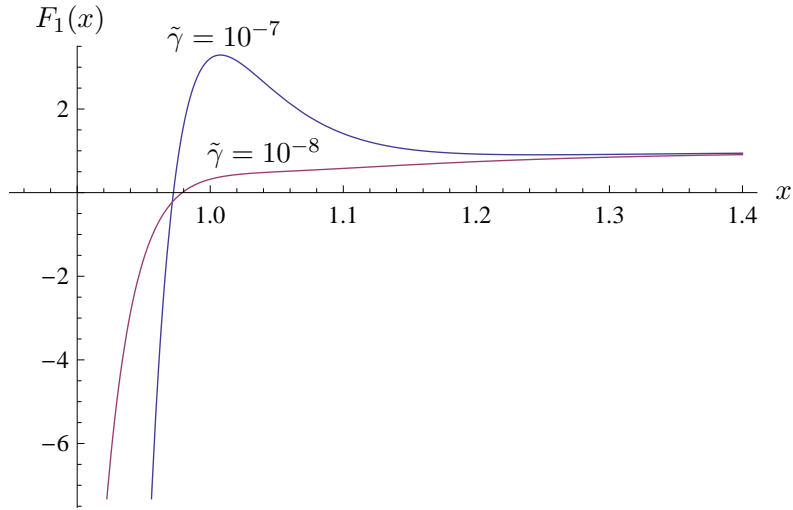}
\end{picture}
\caption{Plots of $F_1(x)$ with $\tilde{\gamma} = 10^{-8}$ and $10^{-7}$.
} \label{fig:F_1}
\end{center}
\end{figure}

Let us examine the properties of $F_1(x)$ given in section \ref{sec:QBS}.
By taking into account the near horizon limit, we set $c_\text{s} = 3747840$ below, 
so $F_1(x)$ is given by
\begin{alignat}{3}
  F_1(x) &= 1 - \frac{1}{x^7} + \tilde{\gamma} \Big( - \frac{1208170880}{9x^{34}} 
  + \frac{161405664}{x^{27}} 
  + \frac{5738880}{13 x^{20}} \notag
  \\
  &\quad\,
  + \frac{956480}{x^{13}}
  + \frac{3747840}{x^7} 
  + \frac{819840}{x^7} I(x) \Big), \qquad
  \tilde{\gamma} = \frac{\gamma}{r_\text{h}^6}.
\end{alignat}
Plots of $F_1(x)$ with $\tilde{\gamma} = 10^{-8}$ and $10^{-7}$ are shown in fig.~\ref{fig:F_1}.
In both cases, locations of the event horizons are shifted inward compared with the classical case.
Especially the behavior of $F_1(x)$ with $\tilde{\gamma}=10^{-7}$ is quite different around the event horizon,
so a test particle feels a repulsive force\cite{Hyakutake:2013vwa}.

%%%%%%%%%%%%%%%%%%%%%%%%%%%%%%%%%%%%%%%%%%%%%%%%%%%%%%%%%%%%%%%%%%%%%%%%%%%%%%%%%%%%%%%%%%%%%%%
%%%%%%%%%%%%%%%%%%%%%%%%%%%%%%%%%%%%%%%%%%%%%%%%%%%%%%%%%%%%%%%%%%%%%%%%%%%%%%%%%%%%%%%%%%%%%%%

\end{document}